\documentclass[final]{IEEEtran}

\usepackage[cmex10]{amsmath}
\usepackage{amssymb}
\usepackage{amsthm}
\usepackage{array}
\usepackage{graphicx}
\usepackage{dsfont}
\usepackage{array}
\usepackage{cite}
\usepackage{stfloats}
\usepackage{shtabularlines}
\usepackage{multirow}
\usepackage{flushend}
\usepackage{tikz}
\usetikzlibrary{calc}

\newtheoremstyle{example}{\topsep}{\topsep}{}{}{\itshape}{:}{.5em}{\thmname{#1}\thmnumber{ #2}\thmnote{ (#3)}}
\newtheoremstyle{examplecontd}{\topsep}{\topsep}{}{}{\itshape}{:}{.5em}{\thmname{#1}\thmnumber{ #2}\thmnote{ #3}\enspace(Cont'd)}
\theoremstyle{example}
\newtheorem{example}{Example}
\theoremstyle{examplecontd}
\newtheorem*{examplecontd}{Example}
\let\fhexamplecontd\examplecontd
\def\examplecontd#1{\fhexamplecontd[\ref{ex: #1}]}

\theoremstyle{example}
\newtheorem{theorem}{Theorem}

\def\remark{
  \let\go\relax
  \ifvmode\vskip-\lastskip\fi
  \noindent{\it Remark\/.}%
  \enskip\relax\ignorespaces\go}
\def\algorithm[#1]#2{%
  \let\go\relax
  \ifvmode\vskip-\lastskip\fi
  \vspace{0.5em}\noindent{\it Algorithm#1: #2\/.}%
  \enskip\relax\ignorespaces\go}
\newcommand{\bs}[1]{\ensuremath{\boldsymbol{#1}}}
\newcommand{\sm}[1]{\mbox{\scriptsize {#1}}}
\newcommand{\secref}[1]{Section~\ref{sec: #1}}
\newcommand{\figref}[1]{Fig.~\ref{fig: #1}}
\newcommand{\tabref}[1]{Table~\ref{tab: #1}}
\newcommand{\tabsref}[2]{Tables~\ref{tab: #1}~--~\ref{tab: #2}}
\newcommand{\tabsandref}[2]{Tables~\ref{tab: #1}~and~\ref{tab: #2}}
\newcommand{\exref}[1]{Example~\ref{ex: #1}}
\newcommand{\thref}[1]{Theorem~\ref{th: #1}}

\newcommand{\cf}{\textit{cf.~}}
\newcommand{\etc}{\textit{etc.~}}

\def\borderarray#1#2#3#4#5#6{%
\setbox0\hbox{$\begin{array}{#5}#6\end{array}$}
\setlength{\dimen1}{\wd0}\addtolength{\dimen1}{-#3}\addtolength{\dimen1}{-\arraycolsep}
\setlength{\dimen2}{\ht0}\addtolength{\dimen2}{-#4}
\setbox1\hbox{$\left#1\rule{\dimen1}{0pt}\rule{0pt}{\dimen2}\right#2$}
\setbox0\hbox{\raisebox{\dp0}{\box0}\kern-\dimen1\kern-7pt\raisebox{-1.5ex}{\box1}}
\vcenter{\box0}
}

\begin{document}

\title{Searching for Voltage Graph-Based LDPC\\Tailbiting Codes with Large Girth}

\author{
  Irina E. Bocharova, Florian Hug,~\IEEEmembership{Student Member,~IEEE}, Rolf Johannesson,~\IEEEmembership{Fellow,~IEEE},\\Boris D. Kudryashov, and Roman V. Satyukov%
  \thanks{This work was supported in part by the Swedish Research Council by Grant 621-2007-61281.}%
  \thanks{I. E. Bocharova, B. D. Kudryashov, and Roman V. Satyukov are with the Department of Information Systems, St. Petersburg University of Information Technologies, Mechanics and Optics, St. Petersburg 197101, Russia (e-mail: irina@eit.lth.se; boris@eit.lth.se, satyukov@gmail.com).}%
  \thanks{F. Hug and R. Johannesson are with the Department of Electrical and Information Technology, Lund University, SE-22100 Lund, Sweden (e-mail: florian@eit.lth.se; rolf@eit.lth.se).}%
}

\maketitle

\begin{abstract}
  The relation between parity-check matrices of quasi-cyclic (QC) low-density parity-check (LDPC) codes and biadjacency matrices of bipartite graphs supports searching for powerful LDPC block codes. Using the principle of tailbiting, compact representations of bipartite graphs based on convolutional codes can be found.
  
  Bounds on the girth and the minimum distance of LDPC block codes constructed in such a way are discussed. Algorithms for searching iteratively for LDPC block codes with large girth and for determining their minimum distance are presented. Constructions based on all-ones matrices, Steiner Triple Systems, and QC block codes are introduced. Finally, new QC regular LDPC block codes with girth up to $24$ are given.
\end{abstract}


\section{Introduction}\label{sec:1}
Low-density parity-check (LDPC) codes, invented by Gallager \cite{Gallager1963} in the $1960$s, constitute a hot research topic since they are a main competitor to turbo codes \cite{Mackay, Chung2001, Lentmaier2010, Kudekar2010}. Recently, a connection between LDPC codes and codes based on graphs was shown (see, for example, \cite{Schmidt2003, Kim2007, Sullivan2006, Barg2006, WovenHypergraph}), which opens new perspectives in searching for powerful LDPC codes. Moreover, coding theory methods can be applied in describing and searching for graphs better than previously known. For example, in \cite{ISIT2009, PPI} compact representations based on convolutional LDPC codes for famous bipartite graphs such as Heawood's, Tutte's, and Balaban's graphs\cite{Bondy1976} are presented.

Typically, LDPC codes have a minimum distance which is less than that of the best known linear codes, but due to their structure they are suitable for low-complexity iterative decoding, like for example the believe-propagation algorithm. An important parameter determining the efficiency of iterative decoding algorithms for LDPC codes is the \textit{girth}, which determines the number of independent iterations\cite{Gallager1963} and is a parameter of the underlying graph. The minimum distance seems not to play an important role within iterative decoding algorithms, since the error-correcting capabilities of such a suboptimal procedure are often less than those guaranteed by the minimum distance. In fact, it was shown in \cite{Dolecek2009} that the performance of LDPC codes in the high signal-to-noise (SNR) region is predominantly dictated by the structure of the smallest absorbing sets. However, as the size of these absorbing sets is upper-bounded by the minimum distance, LDPC codes with large minimum distance are of particular interest. 

LDPC codes can be characterized as either random/pseudo-random or nonrandom, where nonrandom codes can be subdivided into regular or irregular \cite{Kou2001, Johnson2001, Johnson2001-2, Fossorier2004, Kim2007, Sullivan2006, TannerClass, Tanner2002, Tanner2004, Zhang2010, Milenkovic2006, Esmaeili2010, Smarandache2004, Wang2008, Fan2000, ISIT2009, ISIT2010}, while random/pseudo-random codes are always irregular  \cite{Richardson2001, Hu2005}. A $(J,K)$-regular (nonrandom) LDPC code is determined by a parity-check matrix with exactly $J$ ones in each column and exactly $K$ ones in each row. 

The class of quasi-cyclic (QC) $(J, K)$-regular LDPC codes is a subclass of regular LDPC codes with low encoding complexity. Such codes are most suitable for algebraic design and are commonly constructed based on combinatorial approaches using either finite geometries \cite{Kou2001} or Steiner Triple Systems \cite{Johnson2001, Johnson2001-2}, having girth $g \geq 6$. Amongst other algebraic constructions leading to QC LDPC codes with larger girth we would like to mention \cite{TannerClass}, where a class of QC LDPC codes of rate $R=2/5$ with girth up to $12$ based on subgroups of the multiplicative group of the finite field $\mathbb{F}_{p}$ was obtained. The same method was used for convolutional codes in \cite{Tanner2004}.

Although QC LDPC codes are not asymptotically optimal they can outperform random or pseudorandom LDPC codes (from asymptotically optimal ensembles) for short or moderate block lengths \cite{Fossorier2004}. This motivates searching for good short QC LDPC codes.

The problem of finding QC LDPC codes with large girth and large minimum distance for a wide range of code rates was considered in several papers, for example, \cite{Fossorier2004, Kim2007, Tanner2004, Sullivan2006, Esmaeili2010, Milenkovic2006}. Codes with girth at most $12$ are constructed in \cite{TannerClass, Fossorier2004, Tanner2004, Milenkovic2006}, while \cite{Sullivan2006} gives examples of rather short codes with girth $14$. Codes with girth up to $18$ with $J \geq 3$ are presented in \cite{Esmaeili2010} and it is shown that QC LDPC codes with girth $\geq 14$ and block length between $34,\!000$ and $92,\!000$ outperform random codes of the same block length and rate.

Most of the papers devoted to constructing nonrandom LDPC codes with large girth combine some algebraic techniques and computer search. Commonly these procedures start by choosing a proper base matrix (also called weight or degree matrix) or the corresponding base graph (also called seed graph \cite{Tanner2002} or protograph\cite{Thorpe2004}). The references \cite{Tanner2002, Fossorier2004} are focused on all-ones base matrices, while in \cite{Kim2007, Esmaeili2010} base matrices are constructed from Steiner Triple Systems and integer lattices. In both cases, a system of inequalities with integer coefficients describing all cycles of a given length is obtained and suitable labels or degrees are derived. For example, if we replace all nonzero entries in the base matrix by permutation matrices \cite{Thorpe2004, Fan2000}, circulant matrices \cite{Fossorier2004, TannerClass, Sullivan2006, Wang2008, Esmaeili2010}, or sums of circulant matrices \cite{Smarandache2004}, we obtain the corresponding QC LDPC block codes. On the other hand, if we replace all nonzero entries in the base matrix by either monomials or binomials, we obtain the corresponding (parent) LDPC convolutional codes\cite{Tanner2004, ISIT2010, Smarandache2004}.

Notice that both constructing the inequalities and the labeling require significant computational efforts. Some methods directed towards reducing the computational complexity of these steps can be found in \cite{Milenkovic2006, Wang2008}.

Parameters of the so far shortest QC LDPC block codes with $J=3$ and girth $6,8$, and $10$ found via computer search are presented in \cite{Wang2008}, improving previous results from \cite{Fossorier2004}.

As mentioned earlier, it is important that the constructed QC LDPC block codes have large minimum distance for achieving a suitable upper-bound on their error-correcting performance at high SNR. It is proved in \cite{Mackay} that the minimum distance of QC LDPC codes whose base parity-check matrices are $J \times K$ all-one matrices is upper-bounded by $(J+1)!$. However, considering  base matrices with zeros leads to QC LDPC codes with larger minimum distance. For example, in \cite{Smarandache2004} it is shown that replacing all nonzero entries in the base matrix by sums of circulants and all zero entries by all-zero matrices, increases the minimum distance of the resulting code while preserving its regularity. For LDPC convolutional codes this approach implies that a parity-check matrix contains binomials instead of monomials. The corresponding upper-bound on the minimum distance of such LDPC codes is presented in \cite{Smarandache2004}. A particular case of this upper-bound, valid only for codes with zeros and monomials is derived in \cite{ISIT2009}. 

In \secref{2}, we introduce notations for generator and parity-check matrices of convolutional codes and for their corresponding tailbiting block codes. \secref{3} focuses on bipartite graphs, biadjacency matrices, and their relation to parity-check matrices of LDPC block codes. Our constructions of base and voltage matrices, used when searching for LDPC block codes with large girth, are introduced in \secref{4}. Bounds on the girth and the minimum distance for QC $(J, K)$-regular LDPC block codes are discussed in \secref{5}. New search algorithms for QC LDPC block codes constructed from all-one matrices, Steiner Triple Systems, and QC regular matrices are presented in \secref{6}. Moreover, depending on the desired girth, algorithms of different complexity for constructing the set of inequalities and searching for suitable labelings are described. A new algorithm for computing the minimum distance of QC $(J, K)$-regular LDPC codes is described in \secref{7} and used to compute the minimum distance of our newly found codes. Moreover, we determined the hitherto unknown minimum distance for some of the shortest known LDPC codes given in \cite{Esmaeili2010}. In \secref{8}, we present new examples of $(J, K)$-regular QC LDPC codes in the form of tailbiting LDPC codes with girth between $10$ and $24$. This representation is compact and it is possible to apply low-complexity encoding, searching, and decoding procedures well developed for convolutional and tailbiting block codes \cite{RolfKam, beast}. In particular, the presented codes with girth $10$ and $12$ are shorter than the codes presented in \cite{Wang2008} and \cite{Sullivan2006, Zhang2010}, respectively. Moreover, our codes with girth $14$ to $18$ are shorter than the corresponding codes presented in \cite{Esmaeili2010}. \secref{9} concludes the paper with some final remarks.

\section{Generator and Parity-Check Matrices}\label{sec: 2}
Consider a rate $R=b/c$ binary convolutional code $\mathcal{C}$ with the semi-infinite generator matrix
\begin{IEEEeqnarray}{rCl}
  \label{eq: semi-infinite-generator-matrix}
  G & = & \left(\begin{array}{ccccccc}
    G_0 & G_1 & \ldots &        & G_{m_g} \\
        & G_0 & G_1    & \ldots &         & G_{m_g} \\
        &     & \ddots & \ddots &         &         & \ddots
  \end{array}\right)
\end{IEEEeqnarray}
of memory $m_g$ where $G_i$, $i=0,1,\ldots,m_g$, are $b \times c$ binary matrices. Its semi-infinite syndrome former
\begin{IEEEeqnarray}{rCl}
  H^T & = & \left(\begin{array}{ccccccc}
    H^T_0 & H^T_1 & \ldots &        & H^T_m \\
          & H^T_0 & H^T_1  & \ldots &        & H^T_m \\
          &       & \ddots & \ddots &        &        & \ddots
  \end{array}\right)
\end{IEEEeqnarray} 
of memory $m$, where in general $m \neq m_g$, $H_j$, $j=0,1,\ldots,m$, are $(c-b) \times c$ binary matrices, and $T$ denotes transpose. Clearly $G$ and $H$ satisfy
\begin{IEEEeqnarray}{rCl}
  \label{eq: GH-equal-zero}
  G H^T & = & \bs{0}
\end{IEEEeqnarray}
and 
\begin{IEEEeqnarray}{rCl}
  \bs{v} H^T & = & \bs{0}
\end{IEEEeqnarray}
where
\begin{IEEEeqnarray}{rCl}
  \bs{v} & = & \bs{u} G
\end{IEEEeqnarray}
is the code sequence and $\bs{u}$ is the information sequence.

Next we tailbite the semi-infinite generator matrix \eqref{eq: semi-infinite-generator-matrix} to length $M$ $c$-tuples, where $M> \max \{m, m_g\}$. Then we obtain the $Mb \times Mc$ generator matrix of the quasi-cyclic (QC) block code $\mathcal{B}$ as
{\setlength{\arraycolsep}{2pt}
\begin{IEEEeqnarray}{rCl}
  \label{eq: tb-generator-matrix}
  G_{\text{TB}} & = & \left(\begin{array}{ccccccccc}
    G_0         & G_1     & \ldots &         & G_{m_g} \\
                & G_0     & G_1    & \ldots  &         & G_{m_g} \\
                &         & \ddots & \ddots  &         &         & \ddots \\
                &         &        & G_0     & G_1     & \ldots  &        & G_{m_g} \\
    G_{m_g}     &         &        &         & G_0     & G_1     & \ldots & G_{m_g - 1} \\
    G_{m_g - 1} & G_{m_g} &        &         &         & \ddots  & \ddots & \vdots \\
    \vdots      &         & \ddots &         &         &         & \ddots & G_1 \\
    G_1         & G_2     & \ldots & G_{m_g} &         &         &        & G_0
  \end{array}\right)\IEEEeqnarraynumspace
\end{IEEEeqnarray}}
Every cyclic shift of a codeword of $\mathcal{B}$ by $c$ places modulo $Mc$ is a codeword.
The corresponding tailbiting parity-check matrix is the $M(c-b) \times Mc$ matrix
{\setlength{\arraycolsep}{2pt}
\begin{IEEEeqnarray}{rCl}
  \label{eq: tb-parity-check-matrix}
  H_{\text{TB}} & = & \left(\begin{array}{ccccccccc}
    H_0    &        &        &        & H_m     & H_{m-1} & \ldots & H_1 \\
    H_1    & H_0    &        &        &         & H_m     &        & H_2 \\
    \vdots & H_1    & \ddots &        &         &         & \ddots & \vdots \\
           & \vdots & \ddots & H_0    &         &         &        & H_m \\
    H_m    &        &        & H_1    & H_0     &         &        & \\
           & H_m    &        & \vdots & H_1     & \ddots  &        & \\
           &        & \ddots &        & \vdots  & \ddots  & \ddots & \\
           &        &        & H_m    & H_{m-1} & \ldots  & H_1    & H_0
  \end{array}\right)\IEEEeqnarraynumspace
\end{IEEEeqnarray}}
It is easily shown that $G_{\text{TB}}$ and $H_{\text{TB}}$ satisfy
\begin{IEEEeqnarray}{rCl}
  G_{\text{TB}} H_{\text{TB}}^T & = & \bs{0}
\end{IEEEeqnarray}
given that \eqref{eq: GH-equal-zero} is  fulfilled.

The parity-check matrix for the convolutional code $\mathcal{C}$ can also be written as the $(c-b) \times c$ polynomial matrix
\begin{IEEEeqnarray}{rCl}
  H(D) & = & H_0 + H_1 D + H_2 D^2 + \cdots + H_m D^m
\end{IEEEeqnarray}
or, equivalently, as
{\setlength{\arraycolsep}{3pt}
\begin{IEEEeqnarray}{rCl}
  \label{eq: convolutional-parity-check-matrix}
  H(D) & = & \left(\begin{array}{cccc}
    h_{11}(D)     & h_{12}(D)     & \ldots & h_{1c}(D) \\
    h_{21}(D)     & h_{22}(D)     & \ldots & h_{2c}(D) \\
    \vdots        & \vdots        & \ddots \\
    h_{(c-b)1}(D) & h_{(c-b)2}(D) & \ldots & h_{(c-b)c}(D)
  \end{array}\right)\IEEEeqnarraynumspace
\end{IEEEeqnarray}}
In the sequel we mostly consider parity-check matrices with only monomial entries $h_{ij}(D) = D^{w_{ij}}$ of degree $w_{ij}$, where $w_{ij}$ are nonnegative integers. Clearly, such a parity-check matrix $H(D)$ can be represented by its \textit{degree matrix} $W = \left( w_{ij} \right)$, $i=1,2,\ldots,c-b$ and $j=1,2,\ldots,c$. Note that starting from \secref{8} we will relax the restriction to only monomial entries and also include zero entries.

\begin{example}\label{ex: 1}
  Consider the rate $R=1/4$ convolutional code $\mathcal{C}$ with parity-check matrix
  \begin{IEEEeqnarray}{c}
    \label{eq: example_conv_parity_check}
    H(D) = \left( \begin{array}{cccc}
      1 & 1 & 1 & 1 \\
      1 & 1 & D & D \\
      1 & D & 1 & D
    \end{array}\right)
  \end{IEEEeqnarray}
  whose degree matrix is
  \begin{IEEEeqnarray*}{c}
    W = \left( \begin{array}{cccc}
      0 & 0 & 0 & 0 \\
      0 & 0 & 1 & 1 \\
      0 & 1 & 0 & 1
    \end{array}\right)
  \end{IEEEeqnarray*}
  Tailbiting \eqref{eq: example_conv_parity_check} to length $M=2$, we obtain the tailbitten $6 \times 8$ parity-check matrix of a QC LDPC block code
  \begin{IEEEeqnarray}{rCl}
    \label{eq: example_tb_parity_check}
    H_{\rm TB} & = & \borderarray{(}{)}{2em}{1.5ex}{ccccc|cccc}{
      ~~~~ & \sm{1}&\sm{2}&\sm{3}&\multicolumn{1}{c}{\sm{4}}&\sm{5}&\sm{6}&\sm{7}&\sm{8}\vspace{1mm}\\ 
     \sm{1}& 1 & 1 & 1 & 1 & 0 & 0 & 0 & 0 \\
     \sm{2}& 1 & 1 & 0 & 0 & 0 & 0 & 1 & 1 \\
     \sm{3}& 1 & 0 & 1 & 0 & 0 & 1 & 0 & 1 \\ \cline{2-9}
     \sm{4}& 0 & 0 & 0 & 0 & 1 & 1 & 1 & 1 \\
     \sm{5}& 0 & 0 & 1 & 1 & 1 & 1 & 0 & 0 \\                         
     \sm{6}& 0 & 1 & 0 & 1 & 1 & 0 & 1 & 0 \\
    }
  \end{IEEEeqnarray}
  In particular, every cyclic shift of a codeword by $c=4$ places modulo $Mc=8$ is a codeword.
\end{example}

Due to the restriction to monomial entries in $H(D)$, $H_{\text{TB}}$ is $(J,K)$\textit{-regular}, that is, it has exactly $J$ and $K$ ones in each column and row, respectively. Moreover, to fulfill the \textit{low} density criterion, $M$ has to be much larger than $J$ and $K$, and thus the matrix $H_{\text{TB}}$ is sparse.

Note that the first $c$ columns of $H_{\text{TB}}$ are repeated throughout the whole matrix in a cyclicly shifted manner. By reordering the columns as $1,c+1,2c+1,\ldots,(M-1)c+1$, $2,c+2,2c+2,\ldots,(M-1)c+2$, \etc and the rows as $1,(c-b)+1,2(c-b)+1,\ldots,(M-1)(c-b)+1$, $2,(c-b)+2,2(c-b)+2,\ldots,(M-1)(c-b)+2$, \etc we obtain a parity-check matrix of an equivalent $(J, K)$-regular LDPC block code constructed from circulant matrices
\begin{IEEEeqnarray}{rCl}
  \label{eq: circulant-ldpc_parity-check-matrix}
  H_{\text{C}}
  & = & 
  \left( \begin{array}{llll} 
    I_{w_{11}}     & I_{w_{12}}     & \cdots & I_{w_{1c}} \\
    I_{w_{21}}     & I_{w_{22}}     & \cdots & I_{w_{2c}} \\
    \cdots         & \cdots         & \cdots & \cdots     \\
    I_{w_{(c-b)1}} & I_{w_{(c-b)2}} & \cdots & I_{w_{(c-b)c}} \\
  \end{array}\right)\IEEEeqnarraynumspace
\end{IEEEeqnarray}
where $w_{ij}$ are the entries of the degree matrix $W$ and $I_{w_{ij}}$ denotes an $M \times M$ \textit{circulant matrix}, that is, an identity matrix with its rows shifted cyclically to the left by $w_{ij}$ positions. Note, that the $(J, K)$-regular LDPC block code determined by $H_{\text{C}}$ is not quasi-cyclic, although equivalent to the QC block code determined by $H_{\text{TB}}$.

\begin{examplecontd}{1}
  We return to \eqref{eq: example_tb_parity_check} in \exref{1} and reorder the columns as $1,5,2,6,3,7,4,8$ and the rows as $1,4,2,5,3,6$. Then we obtain the equivalent rate $R=1-6/8$ $(3,4)$-regular LDPC block code with parity-check matrix
  \begin{IEEEeqnarray}{rl}
    \label{eq: example_tb_qc_parity_check}
    {\setlength{\arraycolsep}{-3pt}
    \begin{array}{c}
      \\ H_{\text{C}} = 
    \end{array}} & 
     \borderarray{(}{)}{2em}{1.5ex}{ccc|cc|cc|cc}{
      ~~~~ & \sm{1}&\multicolumn{1}{c}{\sm{5}}&\sm{2}&\multicolumn{1}{c}{\sm{6}}&\sm{3}&\multicolumn{1}{c}{\sm{7}}&\sm{4}&\sm{8}\vspace{1mm}\\ 
     \sm{1}& 1 & 0 & 1 & 0 & 1 & 0 & 1 & 0 \\
     \sm{4}& 0 & 1 & 0 & 1 & 0 & 1 & 0 & 1 \\\cline{2-9}
     \sm{2}& 1 & 0 & 1 & 0 & 0 & 1 & 0 & 1 \\
     \sm{5}& 0 & 1 & 0 & 1 & 1 & 0 & 1 & 0 \\\cline{2-9}
     \sm{3}& 1 & 0 & 0 & 1 & 1 & 0 & 0 & 1 \\                         
     \sm{6}& 0 & 1 & 1 & 0 & 0 & 1 & 1 & 0 \\
    }
  \end{IEEEeqnarray}
\end{examplecontd}

\section{Graphs and Biadjacency Matrices}\label{sec: 3}
A graph $\mathcal{G}$ is determined by a set of \textit{vertices} $\mathcal{V} = \{v_i\}$ and a set of \textit{edges} $\mathcal{E} = \{e_i\}$, where each edge connects exactly two vertices. The \textit{degree of a vertex} denotes the number of edges that are connected to it. If all vertices have the same degree $c$, the \textit{degree of the graph} is $c$, or, in other words, the graph is $c$-\textit{regular}.

Consider the set of vertices $\mathcal{V}$ of a graph partitioned into $t$ disjoint subsets $\mathcal{V}_k$, $k=0,1,\ldots,t-1$. Such a graph is said to be $t$-\textit{partite}, if no edge connects two vertices from the same set $\mathcal{V}_k$, $k=0,1,\ldots,t-1$.

A \textit{path} of length $L$ in a graph is an alternating sequence of $L+1$ vertices $v_i$, $i=1,2,\ldots,L+1$, and $L$ edges $e_i$, $i=1,2,\ldots,L$, with $e_i \neq e_{i+1}$. If the first and the final vertex coincide, that is, if $v_1 = v_{L+1}$, then we obtain a \textit{cycle}. A cycle is called \textit{simple} if all its vertices and edges are distinct, except for the first and final vertex which coincide. The length of the shortest simple cycle is denoted the \textit{girth} of the graph. In \cite{Fossorier2004} it was proven that the girth of a graph coincide with the minimum distance of the block code, whose parity-check matrix corresponds to the \textit{incidence matrix} of the graph. Moreover, the girth determines the number of independent iterations in belief-propagation decoding \cite{Gallager1963}.

Every parity-check matrix $H$ of a rate $R=k/n$ LDPC block code can be interpreted as the \textit{biadjacency matrix}\cite{Asratian1998} of a bipartite graph, the so-called \textit{Tanner graph} \cite{Tanner1981}, having two disjoint subsets $\mathcal{V}_0$ and $\mathcal{V}_1$ containing $n$ and $n-k$ vertices, respectively. The $n$ vertices in $\mathcal{V}_0$ are called \textit{symbol nodes}, while the $n-k$ vertices in $\mathcal{V}_1$ are called \textit{constraint nodes}. If the underlying LDPC block code is $(J, K)$-regular, the symbol and constraint nodes have degree $J$ and $K$, respectively.

Consider the Tanner graph with the biadjacency matrix $H_{\text{TB}}$, corresponding to a QC $(J, K)$-regular LDPC code, obtained from the parity-check matrix of a tailbiting LDPC block code. Clearly, by letting the tailbiting length $M$ tend to infinity, we obtain a convolutional parity-check matrix $H(D)$ \eqref{eq: convolutional-parity-check-matrix} of the parent convolutional code $\mathcal{C}$. In terms of Tanner graph representations, this procedure corresponds to unwrapping the underlying graph and extending it in the time domain towards infinity. Hereinafter, we will call the girth of this infinite Tanner graph the \textit{free girth} and denote it $g_{\text{free}}$.

\begin{figure}
  \centering
  \begin{tikzpicture}[scale=1.1]
    \foreach \x in {1,2,3,...,8}
      \path coordinate (S\x) at (1.1*\x, 3);
    \foreach \x in {1,2,3,...,6}
      \path coordinate (C\x) at (1.3*\x+0.6, 0);
    \draw (C1) -- (S1); \draw (C1) -- (S2); \draw (C1) -- (S3); \draw (C1) -- (S4);
    \draw (C2) -- (S1); \draw (C2) -- (S2); \draw (C2) -- (S7); \draw (C2) -- (S8);
    \draw (C3) -- (S1); \draw (C3) -- (S3); \draw (C3) -- (S6); \draw (C3) -- (S8);
    \draw (C4) -- (S5); \draw (C4) -- (S6); \draw (C4) -- (S7); \draw (C4) -- (S8);
    \draw (C5) -- (S3); \draw (C5) -- (S4); \draw (C5) -- (S5); \draw (C5) -- (S6);
    \draw (C6) -- (S2); \draw (C6) -- (S4); \draw (C6) -- (S5); \draw (C6) -- (S7);
    \foreach \x in {1,2,3,...,8}
    {
      \draw[fill=black] (S\x) circle (3pt);
      \node at ($(S\x) + (0,0.3)$) {\makebox(0,0)[b]{\small $s_\x$}};
    }
    \foreach \x in {1,2,3,...,6}
    {
      \draw[fill=white] (C\x) circle (3pt);
      \node at ($(C\x) - (0,0.3)$) {\makebox(0,0)[t]{\small $c_\x$}};
    }
  \end{tikzpicture}
  \caption{\label{fig: example_graph}Tanner graph with $8$ symbol nodes ($s_i$, $i=1,2,\ldots,8$) and $6$ constraint nodes ($c_i$, $i=1,2,\ldots,6$).}
  \vspace{1em}
\end{figure}
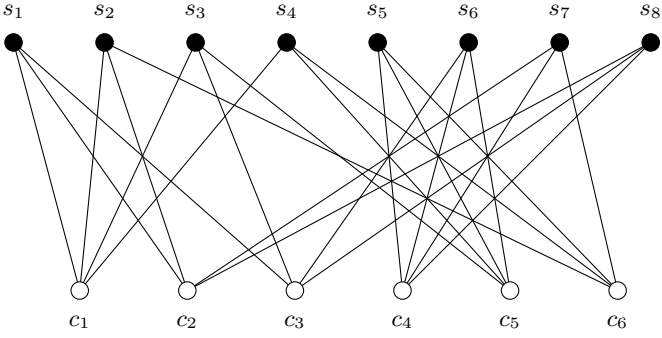

\begin{examplecontd}{1}
  Interpreting \eqref{eq: example_tb_qc_parity_check} as a biadjacency matrix, we obtain the corresponding Tanner graph $\mathcal{G}$ as illustrated in \figref{example_graph} with $8$ symbol nodes and $6$ constraint nodes, having girth $g=4$.
\end{examplecontd}

\section{Base Matrices, Voltages, and their Graphs}\label{sec: 4}
A binary matrix $B$ is called \textit{base matrix} for a tailbiting LDPC block code if its parent convolutional code with parity-check matrix $H(D)$ has only monomial or zero entries and satisfies
\begin{IEEEeqnarray}{c}
  B = H(D)\big|_{D=1}
\end{IEEEeqnarray}
that is, all nonzero entries in $H(D)$ are replaced by $D^0 = 1$. Different tailbiting LDPC block codes can have the same base matrix $B$.

The \textit{base graph} $\mathcal{G}_{\text{B}}$ follows as the bipartite graph, whose biadjacency matrix is given by the base matrix $B$. Denote the girth of such a base graph $g_{\text{B}}$. The terminology ``base graph'' originates from graph theory and is used, for example, in \cite{Kelley2008}. It differs from the terminology used in \cite{Thorpe2004, Sullivan2006}, where \textit{protograph} or \textit{seed graph} are used.

Consider the additive group $(\Gamma, +)$, where $\Gamma=\{\gamma\}$. From the base graph $\mathcal{G}_{\text{B}} = \{\mathcal{E}_{\text{B}},\mathcal{V}_{\text{B}}\}$ we obtain the \emph{voltage graph} \cite{Gross1974,Exoo2008} $\mathcal{G}_{\text{V}} = \{\mathcal{E}_{\text{B}},\mathcal{V}_{\text{B}}, \Gamma\}$ by assigning a voltage value $\gamma(e,v,v^\prime)$ to the edge $e$ connecting the vertices $v$ and $v^\prime$, satisfying the property $\gamma(e,v,v^\prime)=-\gamma(e,v^\prime,v)$. Although the graph is not directed, the voltage of the edge depends on the direction in which the edge is passed. Finally, define the \emph{voltage of the path} to be the sum of the voltages of its edges. 
 
Let $\mathcal{G} = \{\mathcal{E},\mathcal{V}\}$ be a \emph{lifted graph} obtained from a voltage graph $\mathcal{G}_{\text{V}}$, where $\mathcal{E}=\mathcal{E}_{\text{B}}\times \Gamma$ and $\mathcal{V}=\mathcal{V}_{\text{B}}\times \Gamma$. Two vertices $(v,\gamma)$ and $(v^\prime,\gamma^\prime)$ are connected in the lifted graph by an edge if and only if $v$ and $v^\prime$ are connected in the voltage graph $\mathcal{G}_{\text{V}}$ with the voltage value of the corresponding edge given by $\gamma(e,v,v^\prime) = \gamma - \gamma^\prime$.
It is easy to see that cycles in the lifted graph correspond to cycles in the voltage graph with zero voltage. Consequently, the girth $g_{\text{V}}$ of a voltage graph follows as the length of its shortest cycle with voltage zero, which is equal to the free girth $g_{\text{free}}$ \cite{Sullivan2006, Milenkovic2006}. A voltage assignment corresponds directly to selecting the degrees of the parity-check monomials in $H(D)$.

In the following we start from a base graph $\mathcal{G}_{\text{B}}$ and use a voltage assignment based on the monomial degrees $w_{ij}$ of the degree matrix $W$ to determine the corresponding voltage graph $\mathcal{G}_{\text{V}}$. The edge voltage from the constraint node $c_i$ to the symbol node $s_j$ is denoted by $\mu_{ij}$ while the opposite direction from symbol node $s_j$ to constraint node $c_i$ follows as $\bar{\mu}_{ji}$, $i=1,2,\ldots,(c-b)$ and $j=1,2,\ldots,c$, where
\begin{IEEEeqnarray}{C}
  \label{eq: voltage_labeling}
  \left\{ \,
    \begin{IEEEeqnarraybox}[\IEEEeqnarraystrutmode\IEEEeqnarraystrutsizeadd{2pt}{2pt}][c]{rCl}
      \mu_{i j}       & = & w_{ij} \\
      \bar{\mu}_{j i} & = & -w_{ij}
  \end{IEEEeqnarraybox} \right.
\end{IEEEeqnarray}

When searching for LDPC convolutional codes with given free girth $g_{\text{free}}$, we use integer edge voltages, that is, we deal with an infinite additive group. However, when searching for QC LDPC block codes with given girth $g$, obtained by tailbiting a parent convolutional code to length $M$, we use a group of modulo $M$ residues, that is, \eqref{eq: voltage_labeling} is replaced by
\begin{IEEEeqnarray}{C}
  \label{eq: voltage_labeling_mod_M}
  \left\{ \,
    \begin{IEEEeqnarraybox}[\IEEEeqnarraystrutmode\IEEEeqnarraystrutsizeadd{2pt}{2pt}][c]{rCll}
      \mu_{i j}       & = & w_{ij} & \mod M\\
      \bar{\mu}_{j i} & = & -w_{ij} & \mod M
  \end{IEEEeqnarraybox} \right.
\end{IEEEeqnarray}
The definitions of path and cycle in a voltage graph coincide with those in a regular graph, except for the additional restriction that two neighboring edges may not connect the same nodes in reversed order. The \textit{voltage} of a path or cycle within a voltage graph, follows as the sum of all edge voltages involved. 

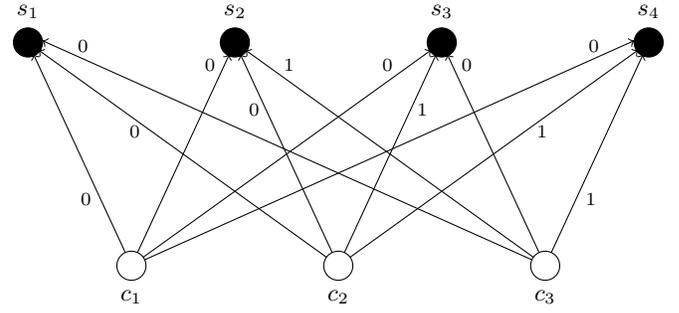
\begin{figure}
  \centering
  \begin{tikzpicture}[scale=1.1]
    \foreach \x in {1,2,3,4}
      \path coordinate (S\x) at (2.5*\x, 2.7);
    \foreach \x in {1,2,3}
      \path coordinate (C\x) at (2.5*\x+1.25, 0);
    \draw[->, shorten >=5pt] (C1) -- (S1) [xshift=-5pt] node[pos=0.3]{\scriptsize $0$}; 
    \draw[->, shorten >=5pt] (C1) -- (S2) [xshift=-5pt] node[pos=0.9]{\scriptsize $0$};
    \draw[->, shorten >=5pt] (C1) -- (S3) [xshift=-8pt] node[pos=0.9]{\scriptsize $0$};
    \draw[->, shorten >=6pt] (C1) -- ($(S4) + (0,0.1)$) [xshift=-10pt] node[pos=0.95]{\scriptsize $0$};
                   
    \draw[->, shorten >=5pt] (C2) -- (S1) [xshift=-6pt] node[pos=0.6]{\scriptsize $0$}; 
    \draw[->, shorten >=5pt] (C2) -- (S2) [xshift=-4pt] node[pos=0.7]{\scriptsize $0$}; 
    \draw[->, shorten >=5pt] (C2) -- (S3) [xshift=+4pt] node[pos=0.7]{\scriptsize $1$}; 
    \draw[->, shorten >=5pt] (C2) -- (S4) [xshift=+6pt] node[pos=0.6]{\scriptsize $1$};
                   
    \draw[->, shorten >=6pt] (C3) -- ($(S1) + (0,0.1)$) [xshift=+10pt] node[pos=0.95]{\scriptsize $0$};
    \draw[->, shorten >=5pt] (C3) -- (S2) [xshift=+8pt] node[pos=0.9]{\scriptsize $1$};
    \draw[->, shorten >=5pt] (C3) -- (S3) [xshift=+5pt] node[pos=0.9]{\scriptsize $0$};
    \draw[->, shorten >=5pt] (C3) -- (S4) [xshift=+5pt] node[pos=0.3]{\scriptsize $1$};
    \foreach \x in {1,2,3,4}
    {
      \draw[fill=black] (S\x) circle (5pt);
      \node at ($(S\x) + (0,0.3)$) {\makebox(0,0)[b]{\small $s_\x$}};
    }
    \foreach \x in {1,2,3}
    {
      \draw[fill=white] (C\x) circle (5pt);
      \node at ($(C\x) - (0,0.3)$) {\makebox(0,0)[t]{\small $c_\x$}};
    }
  \end{tikzpicture}
  \vspace{1em}
  \caption{\label{fig: voltage_graph} Bipartite graph with $4$ symbol nodes ($s_i$, $i=1,2,3,4$) and $3$ constraint nodes ($c_i$, $i=1,2,3$). Since the edges are labeled according to \eqref{eq: voltage_labeling}, this corresponds to a voltage graph.}
\end{figure}

\begin{examplecontd}{1}
  The bipartite graph whose biadjacency matrix is given by the base matrix $B$ of the rate $R=1/4$ $(3, 4)$-regular LDPC convolutional code $\mathcal{C}$ is illustrated in \figref{voltage_graph}. As the edges are labeled according to \eqref{eq: voltage_labeling}, \figref{voltage_graph} corresponds to a voltage graph with girth $g_{\text{V}} = 4$ (for example, $s_1$ $\to$ $c_1$ $\to$ $s_2$ $\to$ $c_2$ $\to$ $s_1$). The edge from, for example, constraint node $c_2$ to symbol node $s_3$ is labeled according to
  \begin{IEEEeqnarray*}{rClClCl}
    \mu_{23} & = & -\bar{\mu}_{32} & = & w_{23} & = & 1 
  \end{IEEEeqnarray*}
  As the free girth of the infinite Tanner graph, corresponding to the parent convolutional code $\mathcal{C}$, determined by the convolutional parity-check matrix $H(D)$ is equal to the girth of the voltage graph, we can conclude that $g_{\text{free}} = g_{\text{V}} = 4$. 
  
  If we neglect all edge labels, we would obtain the corresponding base graph.
\end{examplecontd}

\section{Bounds on the girth and the minimum distance of $(J \geq 3, K)$ QC LDPC block codes}\label{sec: 5}
There are a number of approaches which can be applied to construct and search for QC $(J=2,K)$-regular LDPC block and convolutional codes \cite{PPI} or the bipartite graphs constructed by their incidence matrices. Since every LDPC convolutional code can be represented by a bipartite Tanner graph using the biadjacency matrix, these techniques can be applied to $(J\ge 3,K)$ QC LDPC codes. Moreover, bounds on the girth and the minimum distance of $(J=2,K)$ QC LDPC codes \cite{PPI} can be generalized to an arbitrary $J$.

\begin{theorem}\label{th: one}
  The minimum distance $d_{\text{min}}$ and the girth $g$ of an $(n,k,d_{\text{min}})$ QC LDPC block code $\mathcal{B}$ obtained from a rate $R=b/c$ convolutional code $\mathcal{C}$ with free distance $d_{\text{free}}$ and girth $g_{\rm free}$ by tailbiting to length $M$ are upper-bounded by the inequalities
  \begin{IEEEeqnarray*}{rCl}
    d_{\rm min}     &\leq & d_{\rm free} \\
    g               &\leq & g_{\rm free}
  \end{IEEEeqnarray*}
\end{theorem}
\begin{IEEEproof}
  The first statement follows directly from the fact that any codeword $\bs{v}(D)$ of the tailbiting block code $\mathcal{B}$, obtained from the parity-check matrix $H(D)$ of the parent convolutional code $\mathcal{C}$, satisfies
  \begin{IEEEeqnarray}{c}
    \label{eq: proof_theorem_1}
    \bs{v}(D) H^T(D) = \bs{0} \mbox{ mod } (D^M-1)
  \end{IEEEeqnarray}
  Since the parent convolutional code $\mathcal{C}$ satisfies \eqref{eq: proof_theorem_1} without reduction modulo $(D^M-1)$ and reducing modulo $(D^M-1)$ does not increase the weight of a polynomial, the first statement follows directly.
  
  For the second statement we consider the voltage graph $\mathcal{G}_{\text{V}}$ representation of the parent convolutional code $\mathcal{C}$ with girth $g_{\text{V}} = g_{\text{free}}$ together with the Tanner graph representation of the QC LDPC block code $\mathcal{B}$ with girth $g$. Similar to the relations between the free distance $d_{\text{free}}$ and the minimum distance $d_{\text{min}}$, there exists a relation between each cycle within the voltage graph $\mathcal{G}_{\text{V}}$ of the parent convolutional code and the Tanner graph $\mathcal{G}$ of the corresponding block code obtained by tailbiting to length $M$. The edge voltages for every cycle in $\mathcal{G}_{\text{V}}$ have to sum up to zero. Similarly, every cycle in $\mathcal{G}$ corresponds to a cycle in $\mathcal{G}_{\text{V}}$ such that its edge voltages have to sum up to zero modulo $M$. With the same argument as before it follows directly that
  \begin{IEEEeqnarray*}{rClCl}
    g & \leq & g_{\text{V}}  & = & g_{\text{free}}
  \end{IEEEeqnarray*}
\end{IEEEproof}

In \cite{PPI} a lower bound on the girth of a voltage graph $g_{\text{V}}$ was found via the girth of the corresponding base graph $g_{\text{B}}$ for ordinary graphs. It is straightforward to generalize this bound:

Consider a base graph of a QC $(J \geq 3, K)$-regular LDPC convolutional code with girth $g_{\text{B}}$ and let $d_s$ denote the $s$th generalized minimum Hamming distance of the linear $M((J-2)c+b) \times JMc$ block code determined by the parity-check matrix which corresponds to the incidence matrix of the Tanner graph. In order words, $d_s$ corresponds to the number of nontrivial (not identically zero) positions of an $s$-dimensional linear subcode.

\begin{theorem}\label{th: two}
  There exist a tailbiting length $M$ and a voltage assignment, such that the girth $g$ of the Tanner graph for the corresponding TB block code of length $N=Mc$ satisfies the inequality
  \begin{IEEEeqnarray}{rClCl}
    \label{eq: girth_lower_bound}
    g & \geq & 2 \max \left\{ g_{\text{B}} + \left\lceil g_{\text{B}}/2 \right\rceil, d_2 \right\} & \geq & 3g_{\text{B}}
  \end{IEEEeqnarray}
  where $d_2$ is the second generalized minimum Hamming distance, that is, the minimum support of a subcode of dimension two. We have equality in \eqref{eq: girth_lower_bound}, if the underlying base graph consists of two connected cycles, having at least one common vertex.
\end{theorem}
\begin{IEEEproof}
  According to \thref{one}, any cycle in the Tanner graph of a QC LDPC block code corresponds to a cycle of the same length in the voltage graph. As the labels of the voltage Tanner graph can be freely chosen, it is enough to prove that there is no zero cycle shorter than $2(g_{\text{B}} + \lceil g_{\text{B}}/2 \rceil)$, that is, no such cycle whose voltage is zero regardless of the labeling of the base graph. In particular, such a cycle is also known as an \textit{inevitable cycle} \cite{Kim2007} or \textit{balanced cycle} \cite{Sullivan2006}. The number of times each edge in such a cycle of the voltage graph is passed in different directions has to be even. This cycle cannot be simple, since in a simple cycle each edge is passed in one direction only. Hence, the cycle passes through the vertices of a subgraph which contains at least two different cycles, corresponding to two different nonzero codewords. The minimum distance of the base graph is $g_{\text{B}}$, and, according to the Griesmer bound, the smallest length of a linear code with two nonzero codewords of minimum distance $d$ is $d+ \lceil d/2 \rceil$, and, hence, the first lower bound of inequality \eqref{eq: girth_lower_bound} follows.

  Consider the second lower bound. The definition of the second generalized minimum Hamming distance implies that the smallest subgraph with two cycles has to have at least $d_2$ edges. Thus, the second of the two lower bounds gives the precise value of the girth of a subgraph containing two connected cycles, having at least one common vertex. Otherwise, $d_2$ is a lower bound.
\end{IEEEproof}
The bounds are tighter than the $3g_{\text{B}}$ bound \cite{Kim2007, Kelley2008} but not tight if the shortest non-simple cycle consists of two simple cycles connected by a path.

Finally, we want to recall an upper bound on the achievable girth and minimum distance. We start by reformulating the theorem on the achievable girth by Fossorier \cite{Fossorier2004} and thereby generalize it to include base matrices with zero elements.

\begin{theorem}\label{th: three-a}
  Consider the parity-check matrix $H(D)$ of a rate $R=b/c$ convolutional code with base matrix $B$. Denote the corresponding base graph $\mathcal{G}_{\text{B}}$ and let $B^\prime$ be the $2 \times 3$ submatrix
  \begin{equation}
    B^\prime = \left( \begin{array}{ccc} 1 & 1 & 1 \\ 1 & 1 & 1 \end{array}\right)
  \end{equation}
  If the base matrix $B$, after possibly reordering its rows and columns, contains the submatrix $B^\prime$, then the girth $g_{\text{V}}$ of the corresponding voltage graph $\mathcal{G}_{\text{V}}$ is upper-bounded by
  \begin{equation}
    g_{\text{V}} \leq 12 \label{eq: upper_bound_girth}
  \end{equation}
  regardless of the voltage assignment. 
\end{theorem}  
\begin{IEEEproof}
  The subgraph determined by the $2 \times 3$ submatrix $B^\prime$ contains $3$ symbol nodes, $2$ constraint nodes, and $6$ edges. Moreover, there exist $3$ shortest cycles of length $4$. Thus, the base graph $\mathcal{G}_{\text{B}}$ has girth $g_{\text{B}}=4$ and its second generalized Hamming distance is $d_2 = 6$. Applying \thref{two}, we obtain the precise value of the achievable girth as $2d_2 = 12$, which completes the proof.
\end{IEEEproof}
For parity-check matrices with only nonzero monomial entries, the inequality \eqref{eq: upper_bound_girth} was proven in \cite{Fossorier2004}.

Moreover, let $H(D)$ be the parity-check matrix of a rate $R=b/c$ $(J, K)$-regular LDPC convolutional code with free distance $d_{\text{free}}$. By tailbiting to length $M$ we obtain a QC LDPC block code of block length $Mc$ and minimum distance $d_{\text{min}}$. As proven in \cite{Mackay} for parity-check matrices without zero elements and reformulated in \cite{ISIT2009} for parity-check matrices with zero elements, the corresponding minimum distance $d_{\text{min}}$ can be upper-bounded by
\begin{equation}
  d_{\text{min}} \leq d_{\text{free}} \leq (c-b+1)! \label{eq: upper_bound_minimum_distance}
\end{equation}
For parity-check matrices with only nonzero monomials, the inequality simplifies to $(J+1)!$.

\section{Searching for QC LDPC block codes with large girth}\label{sec: 6}
When searching for QC LDPC block codes with large girth, we start from a base graph of a rate $R=b/c$ $(J, K)$-regular LDPC convolutional code. Using the following algorithm, we determine a suitable voltage assignment based on the group of nonnegative integers, such that the girth of this voltage graph is greater than or equal to a given girth $g$. Afterwards we replace all edge labels by their corresponding modulo $M$ residuals, where we try to minimize $M$ while preserving the girth $g$. Using the concept of biadjacency matrices leads to the corresponding degree matrix $W$ and we obtain the parity-check matrix of a convolutional code whose bipartite graph has girth $g = g_{\text{free}}$. Tailbiting to lengths $M$, yields the rate $R=Mb/Mc$ QC LDPC block code whose parity-check matrix is equal to the biadjacency matrix of a bipartite graph with girth $g$.

The algorithm for determining a suitable voltage assignment for a base graph consists of the following two main steps:
\begin{enumerate}
  \item Construct a list containing all inequalities describing cycles of length smaller than $g$ within the base graph.
  \item Search for such a voltage assignment of the base graph that all inequalities are satisfied.
\end{enumerate}
The efficiency of the second step, searching for a suitable voltage assignment, depends on the chosen representation for the list of inequalities determined during the first step. In general, when searching for all cycles of length $g$ roughly $(J-1)^g$ different paths have been considered. However, by using a similar idea as in \cite{beast} when searching for a path within a trellis, we create a tree of maximum depth $g/2$ and search only for identical nodes within the tree and thereby reduce the complexity to roughly $(J-1)^{g/2}$.

\subsection*{Creating a tree structure}
Utilizing the base graph of a rate $R=b/c$ $(J, K)$-regular LDPC convolutional code, with $c$ symbol and $c-b$ constraint nodes, we construct a separate subtree starting with each of the $c$ symbol nodes.

Before describing the algorithm, we have to introduce some notations. A node in the tree will be denoted by $\xi$ and has a unique parent node $\xi^{\text{p}}$. The underlying base graph is bipartite, that is, every node $\xi$ in the tree with $\xi \in \mathcal{V}_i$ is only connected to nodes $\xi^\prime \in \mathcal{V}_{j}$ with $i,j \in \{0,1\}$, $i \neq j$, where $\mathcal{V}_{0}$ and $\mathcal{V}_{1}$ are the sets of symbol and constraint nodes, respectively. In other words, a symbol node is only connected to constraint nodes and vice versa. Moreover, every node $\xi$ is characterized by its depth $\ell(\xi)$ and its number $n(\xi)$, where $n(\xi) = i$ follows directly from $\xi = s_i$ or $\xi = c_i$ depending on whether its depth $\ell(\xi)$ is even or odd. In particular, $\xi \in \mathcal{V}_{\ell(\xi) \mod 2}$.

Having introduced those basic notations, we can grow $c$ separate subtrees, with the root node $\xi_{i,\text{root}}$ of the $i$th subtree being initialized with $\xi \in \mathcal{V}_0$ and depth $\ell(\xi) = 0$. Extend every node $\xi \in \mathcal{V}_i$ at depth $\ell(\xi) = n$ with $i = n \mod 2$ by connecting it with the nodes $\xi^\prime \in \mathcal{V}_{i+1 \mod 2}$ at depth $n + 1$ according to the underlying base graph, except $\xi^\text{p}$ which is already connected to $\xi$ at depth $n - 1$. 

Finally we label the edges according to \eqref{eq: voltage_labeling} and obtain the voltage for node $\xi$ in the $i$th subtree as the sum of the edge voltages of the path $\xi_{i,\text{root}} \to \xi$.

\begin{figure}
  \centering
  \begin{tikzpicture}
    \foreach \x in {1,2,3,4}
      \path coordinate (S\x) at (2.5*\x, 2.7);
    \foreach \x in {1,2,3}
      \path coordinate (C\x) at (2.5*\x+1.25, 0);
    \draw[->, shorten >=5pt] (C1) -- (S1) [xshift=-10pt] node[pos=0.3]{\scriptsize $\mu_{1 1}$}; 
    \draw[->, shorten >=5pt] (C1) -- (S2) [xshift=-10pt] node[pos=0.9]{\scriptsize $\mu_{1 2}$};
    \draw[->, shorten >=5pt] (C1) -- (S3) [xshift=-13pt] node[pos=0.9]{\scriptsize $\mu_{1 3}$};
    \draw[->, shorten >=6pt] (C1) -- ($(S4) + (0,0.1)$) [xshift=-15pt] node[pos=0.95]{\scriptsize $\mu_{1 4}$};
                   
    \draw[->, shorten >=5pt] (C2) -- (S1) [xshift=-10pt] node[pos=0.6]{\scriptsize $\mu_{2 1}$}; 
    \draw[->, shorten >=5pt] (C2) -- (S2) [xshift=-9pt] node[pos=0.7]{\scriptsize $\mu_{2 2}$}; 
    \draw[->, shorten >=5pt] (C2) -- (S3) [xshift=+11pt] node[pos=0.7]{\scriptsize $\mu_{2 3}$}; 
    \draw[->, shorten >=5pt] (C2) -- (S4) [xshift=+13pt] node[pos=0.6]{\scriptsize $\mu_{2 4}$};
                   
    \draw[->, shorten >=6pt] (C3) -- ($(S1) + (0,0.1)$) [xshift=+14pt] node[pos=0.95]{\scriptsize $\mu_{3 1}$};
    \draw[->, shorten >=5pt] (C3) -- (S2) [xshift=+12pt] node[pos=0.9]{\scriptsize $\mu_{3 2}$};
    \draw[->, shorten >=5pt] (C3) -- (S3) [xshift=+11pt] node[pos=0.9]{\scriptsize $\mu_{3 3}$};
    \draw[->, shorten >=5pt] (C3) -- (S4) [xshift=+11pt] node[pos=0.3]{\scriptsize $\mu_{3 4}$};
    \foreach \x in {1,2,3,4}
    {
      \draw[fill=black] (S\x) circle (5pt);
      \node at ($(S\x) + (0,0.3)$) {\makebox(0,0)[b]{\small $s_\x$}};
    }
    \foreach \x in {1,2,3}
    {
      \draw[fill=white] (C\x) circle (5pt);
      \node at ($(C\x) - (0,0.3)$) {\makebox(0,0)[t]{\small $c_\x$}};
    }
  \end{tikzpicture}
  \caption{\label{fig: voltage_graph_variables} A bipartite voltage graph with $4$ symbol nodes ($s_i$, $i=1,2,3,8$) and $3$ constraint nodes ($c_i$, $i=1,2,3$) with its edges labeled according to \eqref{eq: voltage_labeling}.}
\end{figure}
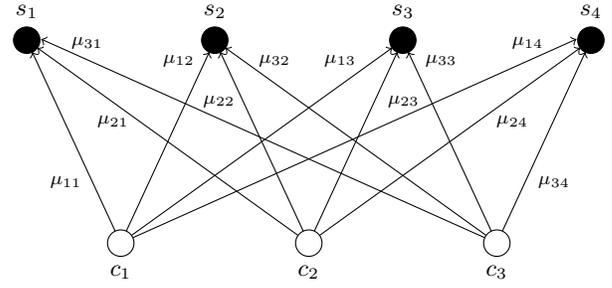

\begin{figure}
  \centering
  \begin{tikzpicture}[scale=1.2, transform shape]
    \path coordinate (start) at (0, 5.6);
    \draw[fill=black] (start) circle (5pt);
    \node[below] at ($(start) - (0,0.15)$) {\small $s_1$};
    \foreach \x in {1,2,3}
    {
      \path coordinate (C\x) at (2.8, 2.8*\x);
      \foreach \y in {2,3,4}
        \path coordinate (S\x\y) at (5, 2.8*\x-2.4+0.8*\y);
    }

    \draw[->, shorten >=6pt] (start) -- (C1) [xshift=+14pt] node[pos=0.4]{\scriptsize $-\mu_{1 1}$};
    \draw[->, shorten >=6pt] (start) -- (C2) [yshift=+5pt, xshift=+14pt] node[pos=0.4]{\scriptsize $-\mu_{2 1}$};
    \draw[->, shorten >=6pt] (start) -- (C3) [xshift=+14pt] node[pos=0.4]{\scriptsize $-\mu_{3 1}$};

    \draw[->, shorten >=6pt] (C1) -- (S12) [yshift=-5pt] node[pos=0.4]{\scriptsize $\mu_{1 2}$};
    \draw[->, shorten >=6pt] (C1) -- (S13) [yshift=+5pt] node[pos=0.65]{\scriptsize $\mu_{1 3}$};
    \draw[->, shorten >=6pt] (C1) -- (S14) [yshift=+9pt] node[pos=0.4]{\scriptsize $\mu_{1 4}$}; 

    \draw[->, shorten >=6pt] (C2) -- (S22) [yshift=-5pt] node[pos=0.4]{\scriptsize $\mu_{2 2}$};
    \draw[->, shorten >=6pt] (C2) -- (S23) [yshift=+5pt] node[pos=0.65]{\scriptsize $\mu_{2 3}$};
    \draw[->, shorten >=6pt] (C2) -- (S24) [yshift=+9pt] node[pos=0.4]{\scriptsize $\mu_{2 4}$}; 

    \draw[->, shorten >=6pt] (C3) -- (S32) [yshift=-5pt] node[pos=0.4]{\scriptsize $\mu_{3 2}$};
    \draw[->, shorten >=6pt] (C3) -- (S33) [yshift=+5pt] node[pos=0.65]{\scriptsize $\mu_{3 3}$};
    \draw[->, shorten >=6pt] (C3) -- (S34) [yshift=+9pt] node[pos=0.4]{\scriptsize $\mu_{3 4}$}; 

    \foreach \x in {1,2,3}
    {
      \draw[fill=white] (C\x) circle (5pt);
      \node[below] at ($(C\x) - (0,0.15)$) {\small $c_\x$};
      \foreach \y in {2,3,4}
      {
        \draw[fill=black] (S\x\y) circle (5pt);
        \node[below] at ($(S\x\y) - (0,0.15)$) {\small $s_\y$};
      }
    }
  \end{tikzpicture}
  \caption{\label{fig: tree} A tree representation with maximum depth two, starting with symbol node $s_1$.}
\end{figure}
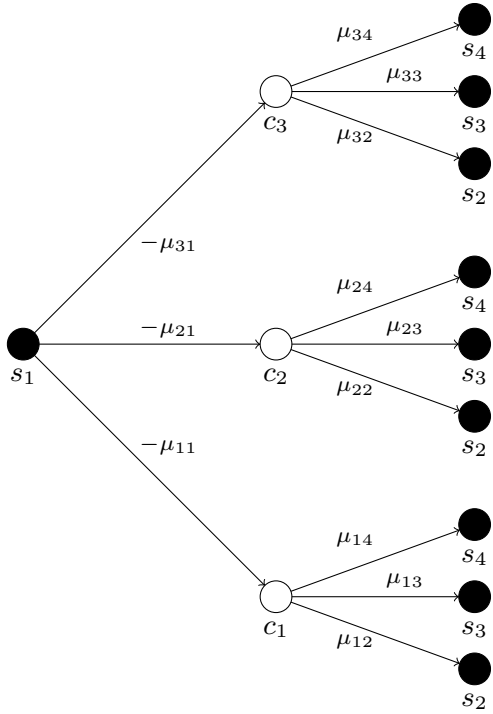

Clearly, all subtrees together contain all paths of a given length in the voltage graph. Moreover, taking into account that the girth $g$ of bipartite graphs is always even, we can conclude that in order to check all possible cycles of length at most $g-2$ in the voltage graph, it is sufficient to grow the corresponding $c$ subtrees up to depth $(g - 2)/2$ and to construct voltage inequalities for all node pairs $(\xi, \xi^\prime)$ in the same subtree $i$ with the same number $n(\xi) = n(\xi^\prime)$ and depth $\ell(\xi) = \ell(\xi^\prime)$ but different parent nodes $\xi^{\text{p}} \neq \xi^{\prime\text{p}}$. 

The corresponding voltage inequality for the node pair $(\xi, \xi^\prime)$ follows directly as the difference between the voltages for the paths from $\xi_{i,\text{root}}$ to $\xi$ and $\xi^\prime$, respectively, that is, $(\xi_{i,\text{root}} \to \xi) - (\xi_{i,\text{root}} \to \xi^\prime)$.

If there exists a cycle $g^\prime < g$, then at depth $g^\prime/2$ there exists at least one pair of nodes $(\xi, \xi^\prime)$, whose corresponding voltage inequality is not satisfied, that is, is equal to zero. If there is no cycle shorter than $g$ in the voltage graph, then there are no such pairs, and all voltage inequalities are satisfied.

\begin{example}\label{ex: 2}
  Consider the rate $R=1/4$ $(3, 4)$-regular LDPC convolutional code given by \eqref{eq: example_conv_parity_check}. The corresponding base graph, with four symbol nodes $s_i \in \mathcal{V}_0$, $i=1,2,3,4$, and three constraint nodes $c_i \in \mathcal{V}_1$, $i=1,2,3$, is illustrated in \figref{voltage_graph_variables}. 
  In the following, we shall search for a set of edge labels, that is, monomial degrees in $W$, such that the corresponding voltage graph has at least a given girth $g$. Thus, we label the branches by the general edge voltages according to \eqref{eq: voltage_labeling} and obtain a bipartite voltage graph.

  In order to find a suitable labeling for the edge voltages from the $i$th constraint node $c_i$ to the $j$th symbol node $s_i$, that is $\mu_{ij}$, $i=1,2,3$, $j=1,2,3,4$, whose underlying voltage graph has at least girth $g = 6$, we have to grow $4$ subtrees up to length $(g-2)/2 = 2$, with their root nodes being initialized by $s_i$, $i=1,2,3,4$.
  
  The subtree with root node $s_1$ is illustrated in \figref{tree}. Clearly, at depth $\ell(\xi)=1$ there are no identical nodes, while at depth $\ell(\xi)=2$ there are $3 \times {3 \choose 2} = 9$ pairs of identical nodes ($n(\xi) = n(\xi^\prime)$), but with different parents. Taking into account that a similar subtree is constructed using the remaining three symbol nodes $s_2$, $s_3$ and $s_4$ as root nodes, we obtain in total $36=4 \times 9$ node pairs, which all correspond to a voltage inequality. 
  
  For example, the voltage inequality obtained by checking all node pairs $(\xi, \xi^\prime)$ with $\xi = s_2$, that is, $n(\xi) = 2$, at depth $\ell(\xi)=2$ in the subtree starting with symbol node $s_1$, are
  \begin{IEEEeqnarray*}{rCl}
    -\mu_{11} + \mu_{12} - \mu_{22} + \mu_{21} & \neq & 0 \\
    -\mu_{11} + \mu_{12} - \mu_{32} + \mu_{31} & \neq & 0 \\
    -\mu_{21} + \mu_{22} - \mu_{32} + \mu_{31} & \neq & 0
  \end{IEEEeqnarray*}
  Note that amongst all $36$ voltage inequalities, there are only $18$ unique ones.
\end{example}

\begin{algorithm}[~TR]{Constructing a tree representation}
  \begin{enumerate}
    \item Grow $c$ separate subtrees according to the underlying base graph up to depth $g/2 -1$, with the root node $\xi_{i,\text{root}}$ of the $i$th subtree being initialized with $\xi \in \mathcal{V}_0$ and depth $\ell(\xi) = 0$.
    \item Extend every node $\xi \in \mathcal{V}_i$ at depth $\ell(\xi) = n$ with $i = n \mod 2$ by connecting it with the nodes $\xi^\prime \in \mathcal{V}_{i+1 \mod 2}$ at depth $n + 1$ according to the underlying base graph, except $\xi^\text{p}$ which is already connected to $\xi$ at depth $n - 1$. Denote the set of all nodes within the $i$th subtree by $\mathcal{T}_i$.
  \end{enumerate}
\end{algorithm}

\subsection*{Searching for a suitable voltage assigment}
Using the obtained subtrees $\mathcal{T}_i$, $i=1,2,\ldots,c$, with maximum depth $g/2 - 1$, we have found all cycles of length smaller than or equal to $g-2$ as well as their corresponding voltage inequalities. 

The same cycle might be found several times within the $c$ subtrees. Moreover, two different cycles can correspond to the same voltage inequality.

We continue by creating a reduced list $\mathcal{L}$ of node pairs $(\xi, \xi^\prime)$ of all $c$ subtrees $\mathcal{T}_i$, $i=1,2,\ldots,c$, containing all unique voltage inequalities. Thereby, we remove all duplicate cycles, as well as different cycles corresponding to the same voltage inequality. Using the reduced list $\mathcal{L}$ we can reduce the obtained $c$ subtrees $\mathcal{T}_i$, $i=1,2,\ldots,c$, in a similar way by removing all nodes, not participating in any of the cycles in $\mathcal{L}$, and denote the reduced subtree by $\mathcal{T}_{i,\text{min}}$. In other words, we remove all nodes in $\mathcal{T}_i$, $i=1,2,\ldots,c$, which only participate in already known cycles or new cycles with already known voltage inequalities.

In the following we present two different approaches for finding suitable edge labels (edge voltages), which we shall denote as Algorithm A and Algorithm B.

In Algorithm A, we label the edges of the reduced subtrees $\mathcal{T}_{i,\text{min}}$, $i=1,2,\ldots,c$, with a set of randomly chosen voltages. For every node pair $(\xi, \xi^\prime)$ in the list $\mathcal{L}$, we calculate the voltage of the corresponding cycle as the difference of the path voltages $\xi_{i,\text{root}} \to \xi$ and $\xi_{i,\text{root}} \to \xi^\prime$. If none of these cycle voltages is equal to zero, the girth of the underlying base graph with such a voltage assignment is greater than or equal to $g$.

In Algorithm B, we discard the list $\mathcal{L}$ and only focus on the $c$ reduced subtrees $\mathcal{T}_{i,\text{min}}$. After labeling their edges with a set of randomly chosen voltages, we sort the nodes $\xi$ of each subtree according to their path voltage $\xi_{i,\text{root}} \to \xi$. If there exists no pair of nodes $(\xi, \xi^\prime)$ with the same path voltage, number $n(\xi)=n(\xi^\prime)$, and depth $\ell(\xi)=\ell(\xi^\prime)$, but different parent nodes $\xi^\text{p} \neq \xi^{\prime\text{p}}$, the girth of the underlying base graph with such a voltage assignment is greater than or equal to $g$.

A formal description of those two algorithms is given below:

\begin{algorithm}[~A]{Constructing a system of voltage inequalities and searching for an optimum voltage assignment using a list}
  \begin{enumerate}
    \item Create a reduced list $\mathcal{L}$ of node pairs $(\xi, \xi^\prime)$ for all $c$ subtrees $\mathcal{T}_i$, $i=1,2,\ldots,c$, containing all node pairs $(\xi, \xi^\prime)$ with a unique voltage inequality, having the same number $n(\xi) = n(\xi^\prime)$, depth $\ell(\xi) = \ell(\xi^\prime)$, but different parent nodes $\xi^{\text{p}} \neq \xi^{\prime\text{p}}$.
    \item Reduce each of the $c$ subtrees $\mathcal{T}_i$ by removing all nodes, which do not participate in any of the found cycles corresponding to the voltage inequalities in $\mathcal{L}$, and denote the reduced subtree structure by $\mathcal{T}_{i,\text{min}}$.
    \item Assign randomly chosen voltages to the edges of all trees and perform the following steps:
    \begin{enumerate}
      \item Find the voltages for all paths leading from the root node $\xi_{i,\text{root}}$ of the $i$th reduced subtree $\mathcal{T}_{i,\text{min}}$ to all nodes $\xi \in \mathcal{T}_{i,\text{min}}$, $i=1,2,\ldots,c$.
      \item Determine the voltage inequality for all cycles $(\xi, \xi^\prime) \in \mathcal{L}$, as the difference of the corresponding path voltages in $\mathcal{T}_{i,\text{min}}$, $i=1,2,\ldots,c$, computed previously.
      \item If all voltage inequalities are satisfied, the girth of the underlying base graph with such a voltage assignment is greater than or equal to $g$.
    \end{enumerate}
  \end{enumerate}
\end{algorithm}

\begin{algorithm}[~B]{Constructing a system of voltage inequalities and searching for an optimum voltage assignment using a tree}
  \begin{enumerate}
    \item Construct the reduced list $\mathcal{L}$ and the reduced subtrees $\mathcal{T}_{i,\text{min}}$, $i=1,2,\ldots,c$, as in Algorithm A without storing the corresponding list $\mathcal{L}$. 
    \item Assign randomly chosen voltages to the edges of all trees and perform the following steps:
    \begin{enumerate}
      \item Find the voltages for all paths from the root node $\xi_{i,\text{root}}$ to all nodes within $\mathcal{T}_{i,\text{min}}$, $i=1,2,\ldots,c$, and sort all elements within $\mathcal{T}_{i,\text{min}}$ according to their voltages.
      \item Search for a pair of nodes $(\xi, \xi^\prime)$ in the sorted list with the same path voltage, number $n(\xi)=n(\xi^\prime)$, and depth $\ell(\xi)=\ell(\xi^\prime)$, but different parent nodes $\xi^\text{p} \neq \xi^{\prime\text{p}}$.
      \item If no such pair exists, then the girth of the corresponding voltage graph with such a voltage assignment is greater than or equal to $g$.
    \end{enumerate}
  \end{enumerate}
\end{algorithm}

\subsection*{Complexity}
Denote the sum of all nodes in the reduced tree $\mathcal{T}_{i,\text{min}}$, $i=1,2,\ldots,c$, and the number of unique inequalities in the list $\mathcal{L}$ by $N_{\text{T}}$ and $N_{\text{L}}$, respectively, that is,
\begin{IEEEeqnarray*}{rCltRCl}
  N_{\text{T}} & = & \sum_{i=1}^c \left| \mathcal{T}_{i,\text{min}}\right| & \quad and \quad &
  N_{\text{L}} & = & \left| \mathcal{L} \right|
\end{IEEEeqnarray*}
where $\left| \mathcal{X} \right|$ denotes the number of entries in the set $\mathcal{X}$.

Algorithm A requires $N_{\text{T}}$ summations for computing the path voltages and $N_{\text{L}}$ comparisons for finding cycles, leading to the complexity estimate $N_{\text{T}} + N_{\text{L}}$. Algorithm B requires the same number of $N_{\text{T}}$ summations for computing the path voltages, roughly $N_{\text{T}} \log_2 N_{\text{T}}$ operations for sorting the set, and $N_{\text{T}}$ comparisons, leading to a total complexity estimate of $N_{\text{T}} \log_2 N_{\text{T}}$. 

\begin{table}
  \centering
  \caption{\label{tab: complexity}Complexity of a search for suitable voltage assignment for QC LDPC block codes with girth $g \leq 12$}
  \renewcommand\arraystretch{1.5}
  \begin{tabular}[t]{c!{\shvline[1pt]}c|c!{\shvline[1pt]}c|c!{\shvline[1pt]}c|c}
    \multirow{2}{*}{$K$} & \multicolumn{2}{c!{\shvline[1pt]}}{$g=8$} & \multicolumn{2}{c!{\shvline[1pt]}}{$g=10$} & \multicolumn{2}{c}{$g=12$} \\
                         & $N_{\text{T}}$ & $N_{\text{L}}$ & $N_{\text{T}}$ & $N_{\text{L}}$ & $N_{\text{T}}$ & $N_{\text{L}}$ \\ \shhline[1pt]
    $4$  &  $53$ &   $42$ &  $150$ &   $231$ &  $269$ &   $519$ \\ \shhline
    $5$  &  $93$ &   $90$ &  $286$ &   $645$ &  $581$ &  $1905$ \\ \shhline
    $6$  & $142$ &  $165$ &  $485$ &  $1470$ & $1060$ &  $5430$ \\ \shhline
    $7$  & $200$ &  $273$ &  $759$ &  $2919$ & $1742$ & $12999$ \\ \shhline
    $8$  & $267$ &  $420$ & $1120$ &  $5250$ & $2663$ & $27426$ \\ \shhline
    $9$  & $343$ &  $612$ & $1580$ &  $8766$ & $3859$ & $52614$ \\ \shhline
    $10$ & $428$ &  $855$ & $2151$ & $13815$ & $5358$ & $93735$ \\ \shhline
    $11$ & $522$ & $1155$ & $2845$ & $20790$ & $7210$ & $157410$ \\ \shhline
    $12$ & $625$ & $1518$ & $3674$ & $30129$ & $9446$ & $251889$ \\ \shhline[1pt]
  \end{tabular}
\end{table}

In \tabref{complexity} the values of $N_{\text{T}}$ and $N_{\text{L}}$ are given when searching for suitable voltage assignment for a $(J, K)$-regular rate $R=1-J/K$ QC LDPC convolutional codes with $J=3$ and arbitrary $K \geq 4$ and girth $g$ constructed from all-ones base matrices. In this case, up to $g = 10$, Algorithm A is preferable, while when searching for voltage assignment with girth $g \geq 12$, Algorithm B should be used. 

In the general case we have to consider all node pairs, and as $N_{\text{L}}$ is roughly $N^2_{\text{T}}$ we conclude that Algorithm B performs asymptotically better (when $N_{\text{T}} \to \infty$). 

\section{Minimum Distance of QC LDPC Codes}\label{sec: 7}
Usually the girth of the Tanner graph of an QC LDPC block code is considered to be the most important parameter that affects the performance of belief-propagation decoding, as it determines the number of independent iterations \cite{Gallager1963}. Therefore, most research is focused on finding QC LDPC block codes with large girth, while their corresponding minimum distance is mostly unknown.  In \cite{Dolecek2009} it was shown, that the performance of belief-propagation decoding algorithms at high SNRs depends on the structure and the size of the smallest absorbing sets, which however can be upper-bounded by the minimum distance. This is the rationale for computing the minimum distance of the shortest known QC LDPC block codes as well as our search for QC LDPC codes with both large girth and large minimum distance.

Our method of calculating the minimum distance is based on the well-known fact that the minimum distance of a linear block code $\mathcal{B}$ with parity-check matrix $H$ is equal to the minimum number of columns of $H$ which sum up to zero.

\begin{table}[t]
  \centering
  \caption{\label{tab: girth6}Degree matrices for QC LDPC codes with girth $g=6$}
  \renewcommand\arraystretch{1.2}
  \setlength\tabcolsep{0.5em}
  \begin{tabular}[t]{c!{\shvline[1pt]}c!{\shvline[1pt]}c!{\shvline[1pt]}c!{\shvline[1pt]}l}
    $K$ & $(n, k)$ & $d_{\text{min}}$ & $M$ & \multicolumn{1}{c}{$W^\prime$} \\ \shhline[1pt]
    \multicolumn{5}{c}{short codes} \\ \shhline[1pt]
    \multirow{2}{*}{$4$}  & \multirow{2}{*}{$(20,     7)$} & \multirow{2}{*}{$6$}  & \multirow{2}{*}{$5$}   & $1,2,4$ \\
                          &                                &                       &                        & $3,1,2$ \\ \shhline
    \multirow{2}{*}{$5$}  & \multirow{2}{*}{$(25,    12)$} & \multirow{2}{*}{$6$}  & \multirow{2}{*}{$5$}   & $1,2,3,4$ \\
                          &                                &                       &                        & $3,1,4,2$ \\ \shhline
    \multirow{2}{*}{$6$}  & \multirow{2}{*}{$(42,    23)$} & \multirow{2}{*}{$4$}  & \multirow{2}{*}{$7$}   & $1,2,3,4,6$ \\ 
                          &                                &                       &                        & $3,5,2,1,4$ \\ \shhline
    \multirow{2}{*}{$7$}  & \multirow{2}{*}{$(49,    30)$} & \multirow{2}{*}{$4$}  & \multirow{2}{*}{$7$}   & $1,2,3,4,5,6$ \\
                          &                                &                       &                        & $3,5,2,1,6,4$ \\ \shhline
    \multirow{2}{*}{$8$}  & \multirow{2}{*}{$(72,    47)$} & \multirow{2}{*}{$4$}  & \multirow{2}{*}{$9$}   & $1,2,3,4,5,7,8$ \\
                          &                                &                       &                        & $3,6,2,1,8,5,4$ \\ \shhline
    \multirow{2}{*}{$9$}  & \multirow{2}{*}{$(81,    56)$} & \multirow{2}{*}{$4$}  & \multirow{2}{*}{$9$}   & $1,2,3,4,5,6,7,8$ \\
                          &                                &                       &                        & $3,6,2,1,8,7,5,4$ \\ \shhline
    \multirow{2}{*}{$10$} & \multirow{2}{*}{$(110,   79)$} & \multirow{2}{*}{$6$}  & \multirow{2}{*}{$11$}  & $1,2,3,4,5 ,6,8,9,10$ \\
                          &                                &                       &                        & $3,1,7,2,10,9,4,6,5 $    \\ \shhline
    \multirow{2}{*}{$11$} & \multirow{2}{*}{$(121,   90)$} & \multirow{2}{*}{$4$}  & \multirow{2}{*}{$11$}  & $1,2,3,4,5 ,6,7,8,9,10$ \\
                          &                                &                       &                        & $3,1,7,2,10,9,8,4,6,5$     \\ \shhline
    \multirow{2}{*}{$12$} & \multirow{2}{*}{$(156,  119)$} & \multirow{2}{*}{$6$}  & \multirow{2}{*}{$13$}  & $1,2,3,4,5,6 ,7,8 ,10,11,12$ \\
                          &                                &                       &                        & $3,1,8,2,9,12,4,11,5, 7, 6, $      \\ \shhline[1pt]
    \multicolumn{5}{c}{large distance codes} \\ \shhline[1pt]                                              
    \multirow{2}{*}{$4$}  & \multirow{2}{*}{$(92,    25)$} & \multirow{2}{*}{$22$} & \multirow{2}{*}{$23$}  & $1,2,4$  \\
                          &                                &                       &                        & $5,3,12$ \\ \shhline
    \multirow{2}{*}{$5$}  & \multirow{2}{*}{$(245,  100)$} & \multirow{2}{*}{$22$} & \multirow{2}{*}{$49$}  & $1,3,10,14$   \\
                          &                                &                       &                        & $40,31,33,30$ \\ \shhline
    \multirow{2}{*}{$6$}  & \multirow{2}{*}{$(414,  209)$} & \multirow{2}{*}{$22$} & \multirow{2}{*}{$69$}  & $3,4,21,26,67$   \\
                          &                                &                       &                        & $34,15,64,33,44$ \\ \shhline
    \multirow{2}{*}{$7$}  & \multirow{2}{*}{$(763,  438)$} & \multirow{2}{*}{$22$} & \multirow{2}{*}{$109$} & $1, 3,11,15,45,93$ \\
                          &                                &                       &                        & $101, 34,18,9,1,4$ \\ \shhline
    \multirow{2}{*}{$8$}  & \multirow{2}{*}{$(1224, 767)$} & \multirow{2}{*}{$22$} & \multirow{2}{*}{$153$} & $2,10,26,57,89,4,49$ \\
                          &                                &                       &                        & $22,19,5,23,61,90,123$ \\ \shhline[1pt]
  \end{tabular}
\end{table}

Consider the $M(c-b) \times Mc$ parity-check matrix $H_{\text{TB}}$ of the $(J,K)$-regular rate $R=Mb/Mc$ tailbiting block code $\mathcal{B}$ with block-length $N=Mc$ \eqref{eq: tb-parity-check-matrix}. Starting with each of the first $c$ columns of $H_{\text{TB}}$ as a root, we will construct $c$ separate trees, where each node $\xi$ is characterized by its depth $\ell(\xi)$ and partial syndrome state column-vector $\bs{\sigma}(\xi)$.

Initialize the partial syndrome state of the root $\xi_{i,\text{root}}$ of the $i$th tree with column $i$ of the corresponding parity-check matrix, that is, $\bs{\sigma}(\xi_{i,\text{root}}) = \bs{h}_i$, $i=1,2,\ldots,c$. Then grow each tree in such a way, that every branch between any two nodes $\xi$ and $\xi^\prime$ is labeled by a column vector $\bs{h}_j$, $j \neq i$, such that $\bs{\sigma}(\xi^\prime) = \bs{\sigma}(\xi) + \bs{h}_j$, where every branch label on the path $\xi_{i,\text{root}} \to \xi^\prime$ does not occur more than once.

Consider now a certain node $\xi$ with nonzero state $\bs{\sigma}(\xi) = (\bs{\sigma}_1(\xi)\,\bs{\sigma}_2(\xi)\ldots\bs{\sigma}_{(c-b)}(\xi))^T$, where $\bs{\sigma}_j(\xi)$, $j=1,2,\ldots,c-b$ is a $M \times 1$ column vector. If we assume that the $k$th position of the column-vector $\bs{\sigma}(\xi)$ is nonzero, then there are at most $K-1$ columns which can cancel this nonzero position and have not been considered previously. Therefore, every node $\xi$ has at most $K-1$ children nodes per nonzero position.

Such a tree would grow until all possible linear combinations have been found. Therefore, we assume that the minimum distance is restricted by $d_{\text{min}} < t$, that is, the maximum depth of the tree is $t-1$. Consequently, a node $\xi$ at depth $\ell(\xi)$ will not be extended, if the number of nonzero positions in its partial syndrome state column-vector $\bs{\sigma}(\xi)$ exceeds $J(t-\ell(\xi)-1)$, since at most $J$ ones can be canceled by each branch.

By initially reordering the columns of the parity-check matrix $H_{\text{TB}}$ in such a way that each of the $c-b$ nonoverlapping blocks of $M$ rows contains not more than a single one per column, we can strengthen the stopping criterion as follows: A node $\xi$ at depth $\ell(\xi)$ will not be extended, if the number of nonzero positions in each of its partial syndrome state column-vectors $\bs{\sigma}_j(\xi)$, $j=1,2,\ldots,c-b$ exceeds $(t-\ell(\xi)-1)$, since at most $1$ one in each block can be canceled by each branch. In particular, such a reordering of the parity-check matrix $H_{\text{TB}}$ corresponds directly to the parity-check matrix $H_{\text{C}}$ \eqref{eq: circulant-ldpc_parity-check-matrix} of the equivalent $(J,K)$-regular LDPC block code constructed from circulant matrices.

\begin{algorithm}[~MD]{Determine the minimum distance of a rate $R=b/c$ $(J, K)$-regular LDPC block code}
  \begin{enumerate}
    \item Assume a suitable restriction $t$ on the minimum distance $d_{\text{min}} < t$.
    \item Grow $c$ separate trees as follows:
    \begin{enumerate}
      \item Initialize the root node of the $i$th tree by $\bs{\sigma}(\xi_{\text{root},i}) = \bs{h_i}$ with depth $\ell(\xi) = 0$.
      \item Extend all nodes $\xi$ as long as the Hamming weights of their partial syndrome states $w_{\text{H}}(\bs{\sigma}(\xi)) \leq J(t-\ell(\xi)-1)$ (Note, for codes with blocks of $M$ rows containing only a single one, this criterion can be strengthen to $w_{\text{H}}(\bs{\sigma}_j(\xi)) \leq (t-\ell(\xi)-1)$, $j=1,2,\ldots,c-b$).
      \item The minimum distance $d_{\text{min}}$ follows directly as
            \begin{IEEEeqnarray*}{rCl}
              d_{\text{min}} & = & \min_{\xi} \left\{ \ell(\xi) \;\vert\; \bs{\sigma}(\xi) = \bs{0} \right\}
            \end{IEEEeqnarray*}
            If there is no node $\xi$ whose partial syndrome state $\bs{\sigma}(\xi) = \bs{0}$, then the minimum distance is lower-bounded by $d_{\text{min}} > t$.
    \end{enumerate}
  \end{enumerate}
\end{algorithm}

\begin{table}[t]
  \centering
  \caption{\label{tab: girth8}Degree matrices for QC LDPC codes with girth $g=8$}
  \renewcommand\arraystretch{1.2}
  \setlength\tabcolsep{0.5em}
  \begin{tabular}[t]{c!{\shvline[1pt]}c!{\shvline[1pt]}c!{\shvline[1pt]}c!{\shvline[1pt]}l}
    $K$ & $(n, k)$ & $d_{\text{min}}$ & $M$ & \multicolumn{1}{c}{$W^\prime$} \\ \shhline[1pt]
    \multicolumn{5}{c}{short codes} \\ \shhline[1pt]
    \multirow{2}{*}{$4$}  & \multirow{2}{*}{$(36,    11)$}                       & \multirow{2}{*}{$6$}  & \multirow{2}{*}{$9$}   & $1,4,6$ \\
                          &                                                      &                       &                        & $5,2,3$ \\ \shhline
    \multirow{2}{*}{$5$}  & $(65,    28)$                                        & \multirow{2}{*}{$10$} & \multirow{2}{*}{$13$}  & $1,3,7,11$ \\
                          & \scriptsize \big($(75,  32)$\cite{Esmaeili2010}\big) &                       &                        & $10,4,5,6$ \\ \shhline
    \multirow{2}{*}{$6$}  & $(108,   56)$                                        & \multirow{2}{*}{$10$} & \multirow{2}{*}{$18$}  & $2,3,5,7,9$   \\
                          & \scriptsize \big($(156, 80)$\cite{Esmaeili2010}\big) &                       &                        & $4,6,13,1,16$ \\ \shhline
    \multirow{2}{*}{$7$}  & \multirow{2}{*}{$(147,   86)$}                       & \multirow{2}{*}{$10$} & \multirow{2}{*}{$21$}  & $2,3,8,15,17,20$ \\
                          &                                                      &                       &                        & $4,6,7,9,12,13$  \\ \shhline
    \multirow{2}{*}{$8$}  & \multirow{2}{*}{$(200,  127)$}                       & \multirow{2}{*}{$8$}  & \multirow{2}{*}{$25$}  & $1,3,4,10,14,15,19$ \\
                          &                                                      &                       &                        & $5,6,11,24,2,9,12$  \\ \shhline
    \multirow{2}{*}{$9$}  & \multirow{2}{*}{$(270,  182)$}                       & \multirow{2}{*}{$8$}  & \multirow{2}{*}{$30$}  & $1,3,10,16,23,25,26,28$ \\ 
                          &                                                      &                       &                        & $2,6,5,9,8,12,14,22$    \\ \shhline
    \multirow{2}{*}{$10$} & \multirow{2}{*}{$(350,  247)$}                       & \multirow{2}{*}{$8$}  & \multirow{2}{*}{$35$}  & $2,6,7,18,19,26,29,31,34$ \\ 
                          &                                                      &                       &                        & $4,5,3,13,10,16,12,11,23$ \\ \shhline
    \multirow{2}{*}{$11$} & \multirow{2}{*}{$(451,  330)$}                       & \multirow{2}{*}{$8$}  & \multirow{2}{*}{$41$}  & $1,4,8,20,27,28,29,33,39,40$ \\ 
                          &                                                      &                       &                        & $5,7,6,9,10,19,13,21,14,35$  \\ \shhline
    \multirow{2}{*}{$12$} & \multirow{2}{*}{$(564,  425)$}                       & \multirow{2}{*}{$8$}  & \multirow{2}{*}{$47$}  & $3,7,8,22,24,27,29,35,40,41,43$ \\ 
                          &                                                      &                       &                        & $6,2,4,5,14,16,1,21,28,9,34$    \\ \shhline[1pt]
    \multicolumn{5}{c}{large distance codes} \\ \shhline[1pt] 
    \multirow{2}{*}{$4$}  & \multirow{2}{*}{$(116,   31)$}                       & \multirow{2}{*}{$24$} & \multirow{2}{*}{$29$}  & $3,14,21$ \\
                          &                                                      &                       &                        & $7,1,17$  \\ \shhline
    \multirow{2}{*}{$5$}  & \multirow{2}{*}{$(225,   92)$}                       & \multirow{2}{*}{$24$} & \multirow{2}{*}{$45$}  & $1,3,10,14$   \\
                          &                                                      &                       &                        & $40,31,33,30$ \\ \shhline
    \multirow{2}{*}{$6$}  & \multirow{2}{*}{$(431,  218)$}                       & \multirow{2}{*}{$24$} & \multirow{2}{*}{$72$}  & $3,4,21,26,67$   \\
                          &                                                      &                       &                        & $34,15,64,33,44$ \\ \shhline
    \multirow{2}{*}{$7$}  & \multirow{2}{*}{$(777,  446)$}                       & \multirow{2}{*}{$24$} & \multirow{2}{*}{$111$} & $3,11,15,45,93,110$ \\
                          &                                                      &                       &                        & $34,18,9,1,4,101$   \\ \shhline
    \multirow{2}{*}{$8$}  & \multirow{2}{*}{$(1280, 802)$}                       & \multirow{2}{*}{$24$} & \multirow{2}{*}{$160$} & $2,4,10,26,49,57,89$   \\
                          &                                                      &                       &                        & $22,90,19,5,123,23,61$ \\ \shhline
    \multirow{2}{*}{$9$}  & \multirow{2}{*}{$(1386, 926)$}                       & \multirow{2}{*}{$20$} & \multirow{2}{*}{$154$} & $6,9,26,65,79,99,124,153$ \\ 
                          &                                                      &                       &                        & $24,16,14,1,46,62,137,84$ \\ \shhline[1pt]
  \end{tabular}
\end{table}

\section{Search results}\label{sec: 8}
When presenting our search results for QC ($J=3, K$)-regular LDPC block codes with different girth we will distinguish two cases. 

We started by searching for QC ($J=3, K$)-regular LDPC block codes using an all-one base matrix $B$, applied the algorithms as described above and obtained QC ($3, K$)-regular LDPC block codes with girth $g=6,8,10$, and $12$ as given in \tabsref{girth6}{girth12}. These codes correspond to a parity-check matrix $H(D)$ of a convolutional code with only monomial entries as given, for example, in \eqref{eq: convolutional-parity-check-matrix}.

However, according to \thref{three-a} the achievable girth $g$ of a QC $(J,K)$-regular LDPC code, constructed in such a way, is limited by $g \leq 12$. Thus, in order to find QC $(J=3, K)$-regular LDPC block codes with girth $g > 12$ as presented in \tabsandref{girthgreater}{girth20}, we have to allow zero entries in our base matrix $B$. This is a straight-forward generalization of the restriction to only monomial entries in the parity-check matrix $H(D)$ of the underlying convolutional code.

\begin{table}
  \centering
  \caption{\label{tab: girth10}Degree matrices for QC LDPC codes with girth $g=10$}
  \renewcommand\arraystretch{1.2}
  \setlength\tabcolsep{0.4em}
  \begin{tabular}[t]{c!{\shvline[1pt]}c!{\shvline[1pt]}c!{\shvline[1pt]}c!{\shvline[1pt]}l}
    $K$ & $(n, k)$ & $d_{\text{min}}$ & $M$ & \multicolumn{1}{c}{$W^\prime$} \\ \shhline[1pt]
    \multicolumn{5}{c}{short codes} \\ \shhline[1pt]
    \multirow{2}{*}{$4$}  & $(148, 39)$                                           & \multirow{2}{*}{$14$} & $37$                                         & $1,14,17$ \\
                          & \scriptsize \big($(144, 38)$\cite{Esmaeili2010}\big)  &                       & \scriptsize \big($39$\cite{Wang2008}\big)    & $11,6,2$  \\ \shhline
    \multirow{2}{*}{$5$}  & $(305, 124)$                                          & \multirow{2}{*}{$24$} & $61$                                         & $2,20,54,60$  \\
                          & \scriptsize \big($(550, 222)$\cite{Esmaeili2010}\big) &                       & \scriptsize \big($61$\cite{TannerClass}\big) & $26,16,31,48$ \\ \shhline
    \multirow{2}{*}{$6$}  & $(606, 305)$                                          & \multirow{2}{*}{$24$} & $101$                                       & $2,24,25,54,85$ \\
                          & \scriptsize \big($(780, 392)$\cite{Esmaeili2010}\big) &                       & \scriptsize \big($103$\cite{Wang2008}\big)  & $21,15,11,8,59$ \\ \shhline
    \multirow{2}{*}{$7$}  & \multirow{2}{*}{$(1113, 638)$}                        & \multirow{2}{*}{$24$} & $159$                                       & $2,14,27,67,97,130$ \\
                          &                                                       &                       & \scriptsize \big($160$\cite{Wang2008}\big)  & $21,24,1,6,75,58$   \\ \shhline
    \multirow{2}{*}{$8$}  & \multirow{2}{*}{$(1752, 1097)$}                       & \multirow{2}{*}{$24$} & $219$                                       & $3,14,26,63,96,128,183$ \\
                          &                                                       &                       & \scriptsize \big($233$\cite{Wang2008}\big)  & $24,6,19,46,4,77,107$   \\ \shhline
    \multirow{2}{*}{$9$}  & \multirow{2}{*}{$(2871, 1916)$}                       & \multirow{2}{*}{$24$} & $319$                                       & $6,9,26,65,99,153,233,278$ \\
                          &                                                       &                       & \scriptsize \big($329$\cite{Wang2008}\big)  & $24,16,14,1,62,84,200,137$ \\ \shhline
    \multirow{4}{*}{$10$} & \multirow{4}{*}{$(4300, 2912)$}                       & \multirow{4}{*}{$24$} &                                             & $9,11,26,67,101,161,\ldots$ \\
                          &                                                       &                       & $430$                                       & $\quad 233,302,395$ \\
                          &                                                       &                       & \scriptsize \big($439$\cite{Wang2008}\big)  & $23,5,1,54,33,96,\ldots$ \\
                          &                                                       &                       &                                             & $\quad 120,104,244$ \\ \shhline
    \multirow{4}{*}{$11$} & \multirow{4}{*}{$(6160, 4482)$}                       & \multirow{4}{*}{$24$} &                                             & $2,11,25,62,101,162,225,\ldots$ \\
                          &                                                       &                       & $560$                                       & $\quad 268,421,492$ \\
                          &                                                       &                       & \scriptsize \big($577$\cite{Wang2008}\big)  & $24,21,5,55,6,59,178,\ldots$ \\
                          &                                                       &                       &                                             & $\quad 132,204,311$    \\ \shhline
    \multirow{4}{*}{$12$} & \multirow{4}{*}{$(8844, 6635)$}                       &                       &                                             & $2,22,23,63,101,147,219,\ldots$ \\
                          &                                                       &                       & $737$                                       & $\quad 322,412,569,601$ \\
                          &                                                       &                       & \scriptsize \big($758$\cite{Wang2008}\big)  & $16,9,6,58,34,91,126,\ldots$ \\
                          &                                                       &                       &                                             & $\quad 155,185,298,232$ \\ \shhline[1pt]
    \multicolumn{5}{c}{large distance codes} \\ \shhline[1pt] 
    \multirow{2}{*}{$4$}  & \multirow{2}{*}{$(176, 46)$}                          & \multirow{2}{*}{$24$} & \multirow{2}{*}{$44$}                       & $1,14,17$ \\
                          &                                                       &                       &                                             & $11,6,2$  \\ \shhline[1pt]
  \end{tabular}
\end{table}

\begin{table}[t]
  \centering
  \caption{\label{tab: girth12}Degree matrices for QC LDPC codes with girth $g=12$}
  \renewcommand\arraystretch{1.2}
  \setlength\tabcolsep{0.3em}
  \begin{tabular}[t]{c!{\shvline[1pt]}c!{\shvline[1pt]}c!{\shvline[1pt]}c!{\shvline[1pt]}l}
    $K$ & $(n, k)$ & $d_{\text{min}}$ & $M$ & \multicolumn{1}{c}{$W^\prime$} \\ \shhline[1pt]
    \multicolumn{5}{c}{short codes} \\ \shhline[1pt]
    \multirow{2}{*}{$4$}  & $(292, 75)$                                             & \multirow{2}{*}{$24$} & $73$                                            & $2,25,33$ \\
                          & \scriptsize \big($( 444, 113)$\cite{Esmaeili2010}\big)  &                       & \scriptsize \big($97$\cite{Sullivan2006}\big)   & $18,6,5$  \\ \shhline
    \multirow{2}{*}{$5$}  & $(815, 328)$                                            & \multirow{2}{*}{$24$} & $163$                                           & $5,33,42,117$ \\
                          & \scriptsize \big($(1700, 682)$\cite{Esmaeili2010}\big)  &                       & \scriptsize \big($181$\cite{TannerClass}\big)   & $36,35,25,57$ \\ \shhline
    \multirow{2}{*}{$6$}  & $(1860, 932)$                                           & \multirow{2}{*}{$24$} & $310$                                           & $1,24,38,145,246$ \\
                          & \scriptsize \big($(4680, 2342)$\cite{Esmaeili2010}\big) &                       & \scriptsize \big($393$\cite{Zhang2010}\big)     & $16,36,5,82,110$  \\ \shhline
    \multirow{2}{*}{$6$}  & \multirow{2}{*}{$(1836, 920)$}                          & \multirow{2}{*}{$24$} & $306$                                           & $9,36,38,154,204$ \\
                          &                                                         &                       & \scriptsize \big($393$\cite{Zhang2010}\big)     & $33,1,13,54,123$  \\ \shhline
    \multirow{2}{*}{$7$}  & \multirow{2}{*}{$(3962, 2266)$}                         &                       & $566$                                           & $3,10,33,147,297,442$ \\
                          &                                                         &                       & \scriptsize \big($881$\cite{Sullivan2006}\big)  & $31,22,4,93,133,219$  \\ \shhline
    \multirow{2}{*}{$8$}  & \multirow{2}{*}{$(6784, 4242)$}                         &                       & $848$                                           & $4,24,31,143,303,498,652$ \\
                          &                                                         &                       & \scriptsize \big($1493$\cite{Sullivan2006}\big) & $32,9,6,70,130,193,222$   \\ \shhline
    \multirow{4}{*}{$9$}  & \multirow{4}{*}{$(12384, 8258)$}                        &                       &                                                 & $4,20,32,160,284,\ldots$ \\
                          &                                                         &                       & $1376$                                          & $\quad 569,794,1133$     \\
                          &                                                         &                       & \scriptsize \big($2087$\cite{Sullivan2006}\big) & $30,7,1,92,169,\ldots$   \\
                          &                                                         &                       &                                                 & $\quad 350,437,645$ \\ \shhline
    \multirow{4}{*}{$10$} & \multirow{4}{*}{$(21030, 14723)$}                       &                       & \multirow{4}{*}{$2103$}                         & $6,13,28,150,291,565,\ldots$ \\
                          &                                                         &                       &                                                 & $\quad 678,1258,1600$ \\
                          &                                                         &                       &                                                 & $30,16,5,64,225,207,\ldots$ \\
                          &                                                         &                       &                                                 & $\quad 491,838,746$ \\ \shhline
    \multirow{4}{*}{$11$} & \multirow{4}{*}{$(34507, 25098)$}                       &                       & \multirow{4}{*}{$3137$}                         & $9,11,24,150,306,508,\ldots$ \\
                          &                                                         &                       &                                                 & $\quad 666,1279,1765,1958$ \\
                          &                                                         &                       &                                                 & $31,28,1,83,131,160,\ldots$ \\
                          &                                                         &                       &                                                 & $\quad 429,550,956,1391$ \\ \shhline
    \multirow{4}{*}{$12$} & \multirow{4}{*}{$(56760, 42572)$}                       &                       & \multirow{4}{*}{$4730$}                         & $3,15,22,140,286,537,\ldots$ \\
                          &                                                         &                       &                                                 & $\quad 811,1113,1878,2524,3349$ \\
                          &                                                         &                       &                                                 & $31,26,1,66,95,210,373,\ldots$ \\
                          &                                                         &                       &                                                 & $\quad 729,878,1365,1644$ \\ \shhline[1pt]
  \end{tabular}
\end{table}

\subsection*{Case I: monomial entries}
In \tabsref{girth6}{girth12}, parity-check matrices of short known QC $(J=3, K)$-regular LDPC block codes with girth $g=6,8,10$, and $12$ together with those of large minimum distance are presented. When searching for such codes, we applied the following restrictions to reduce the number of possible voltage assignments:
\begin{itemize}
  \item As the girth of a voltage graph is defined as the shortest cycle with voltage zero, and the sign of the voltage depends on the direction of the edge, we can add the same arbitrary offset to the voltage of all edges being connected to the same node. Thus, without loss of generality, we set the voltage of all edges connected to one specific symbol node as well as all edges connected to one specific constraint node to voltage zero. (For consistency with codes constructed from Steiner Triple System, that will be introduced later, we choose the first symbol node and the last constraint node. This corresponds directly to a degree matrix with zeros in its first column and last row.)

	For example, the degree matrix of the $(J=3, K=4)$ QC LDPC block code with girth $g=8$ from \tabref{girth8} follows directly as
	\begin{IEEEeqnarray*}{rCl}
		W & = &
		\left(\begin{array}{cccc}
			0 & 1 & 4 & 6 \\
			0 & 5 & 2 & 3 \\
			0 & 0 & 0 & 0
		\end{array}\right)
	\end{IEEEeqnarray*}
	
  \item Furthermore, to reduce the number of only permuted degree matrices, we assume that
  \begin{itemize}
    \item The first row is sorted in ascending order.
    \item When sorting the first and the second row in ascending order, the second row is lexicographically less than the first row.
    \item The maximum degree is less than the tailbiting length $M$ for which there exists a QC $(J=3, K)$-regular LDPC block code with given girth $g$.
  \end{itemize}
  \item QC $(J=3, K=4)$-regular LDPC block codes were found by exhaustive search over the previously defined set of restricted edge voltages. 
  \item QC $(J=3, K=N)$-regular LDPC block codes with $N > 4$ were obtained by adding one additionally, randomly chosen column to the best degree matrices of codes with $K=N-1$ having the same girth $g$. The maximum degree in this additional column is limited by twice the maximum degree of the previous code.
\end{itemize}
Using these restrictions, the obtained QC $(J=3, K)$-regular LDPC block codes with girth $g=6,8,10$, and $12$ are presented in \tabsref{girth6}{girth12}. 

The first column $K$ denotes the number of nonzero elements per row, which corresponds to the number of columns in $H(D)$ and $W$, due to the all-ones base matrix $B$. As all entries in the first column and the last row of the degree matrix $W$ are zero, they are omitted in the submatrix $W^\prime$ which is given in the corresponding last column. 

Consider now the parity-check matrix $H(D)$ of the rate $R=1-J/K$ convolutional code $\mathcal{C}$, with only monomial entries corresponding to the degree matrix $W$. By tailbiting the semi-infinite parity-check matrix $H$ to length $M$ (given in the forth column), we obtain the parity-check matrix $H_{\text{TB}}$ of an $(n, k)$ block code $\mathcal{B}$ with minimum distance $d_{\text{min}}$, where $(n, k)$ and $d_{\text{min}}$ follow from the second and third column, respectively. Note that due to linear dependent rows in $H_{\text{TB}}$ the rank of $\mathcal{B}$ might be less than $M(c-b)$.

The codes presented in \tabsandref{girth6}{girth8} coincide with the QC LDPC block codes found by the ``hill-climbing'' algorithm \cite{Wang2008}, but we determined their minimum distance with our algorithm described in \secref{7}. \tabsandref{girth10}{girth12} contain new QC $(J=3, K)$-regular LDPC block codes, which, to the best of our knowledge, are shorter than the previously known codes obtained from an all-ones base matrix \cite{Wang2008, TannerClass, Sullivan2006, Zhang2010}. In particular, these codes are significantly shorter than those presented in \cite{Esmaeili2010}, which are obtained from base matrices with zeros. However, due to the zeros in their base matrix, the minimum distance of the LDPC block codes in \cite{Esmaeili2010} can exceed $(J+1)!$. For example, we determined the minimum distance of the $(444, 113)$ QC $(3, 4)$-regular LDPC block code with girth $g=12$ in \cite{Esmaeili2010} to be $d_{\text{min}} = 28$, while the corresponding code in \tabref{girth12}, that is, the $(292, 75)$ QC $(3, 4)$-regular LDPC block code, has only minimum distance $d_{\text{min}}=24$, but shorter block length. Using the BEAST \cite{beast}, we calculated the free distance of the corresponding parent convolutional code for the code in \cite{Esmaeili2010} to be $d_{\text{free}} = 46$. Therefore, using our approach and a larger tailbiting length it would be possible to construct corresponding QC $(3, 4)$-regular LDPC block codes with minimum distance up to $46$.

\subsection*{Case 2: monomial or zero entries}
In order to find QC $(J=3, K)$-regular LDPC block codes with girth $g \geq 14$, we have to allow zero entries in our base matrix $B$; that is, relax the restriction from only monomial entries in $H(D)$ to include also zero entries. According to \thref{two}, a code with girth $g$ exists if the corresponding base graph has girth $g_{\text{B}}$ satisfying \eqref{eq: girth_lower_bound}. Additionally, as we are searching for codes with short block length, we consider the shortest possible base matrices $B$.

\begin{table*}[t]
  \centering
  \caption{\label{tab: girthgreater}Degree matrices for QC LDPC codes with girth $g = 14$ to $18$}
  \renewcommand\arraystretch{1.2}
  \setlength\tabcolsep{0.3em}
  \begin{tabular}[t]{c!{\shvline[1pt]}c!{\shvline[1pt]}c!{\shvline[1pt]}c!{\shvline[1pt]}c!{\shvline[1pt]}l}
    $K$ & $g$ & $(n, k)$ & $M$ & Base graph & \multicolumn{1}{c}{$W^{\prime}$} \\ \shhline[1pt]
    \multirow{2}{*}{$4$} & \multirow{2}{*}{$14$} & $(1812, 453)$                                               & $151$                                           & $\text{STS}(9)$     & $0, 123, 36, 3, 2, 79, 4, 7, 52, 4, 1$ \\
                         &                       & \scriptsize \big($(2208, 732)$\cite{Esmaeili2010}\big)      & \scriptsize \big($184$\cite{Esmaeili2010}\big)  & $(9 \times 12)$     & $0, 96, 23, 11, 1, 37, 12, 2, 61, 1, 4$ \\ \shhline
    \multirow{2}{*}{$5$} & \multirow{2}{*}{$14$} & $(9720, 3888)$                                              & \multirow{2}{*}{$486$}                          & $\text{S-STS}(13)$  & $423, 0, 437, 5, 237, 235, 170, 333, 260, 109, 241, 2, 114, 5, 2, 428, 92, 228, 299$ \\
                         &                       & \scriptsize \big($(11525, 4612)$\cite{Esmaeili2010}\big)    &                                                 & $(12 \times 20)$    & $0, 0, 0, 445, 465, 51, 440, 22, 111, 307, 433, 4, 285, 2, 1, 4, 113, 282, 5$\\ \shhline
    \multirow{4}{*}{$6$} & \multirow{4}{*}{$14$} &                                                             &                                                 &                     & $1037, 0, 1051, 1105, 933, 1027, 962, 1000, 665, 805, 646, 2,  \ldots$ \\
                         &                       & $(29978, 14989)$                                            & $1153$                                          & $\text{STS}(13)$    & $\quad 906, 5, 2, 1095, 788, 633, 913, 264, 51, 772, 672, 686, 737$ \\
                         &                       & \scriptsize \big($(37154, 18579)$\cite{Esmaeili2010}\big)   & \scriptsize \big($1429$\cite{Esmaeili2010}\big) & $(13 \times 26)$    & $0, 0, 0, 1112, 1132, 51, 1107, 22, 807, 921, 1100, 4, 952, 2,  \ldots$ \\
                         &                       &                                                             &                                                 &                     & $\quad 1, 4, 905, 949, 5, 0, 1111, 922, 620, 351, 140$ \\ \shhline
    $7$                  & $14$                  & $n=80000000$                                                & $800000$                                        & $\text{STS}(25)$    & \quad \textit{available at \cite{online}}\\ \shhline
    \multirow{2}{*}{$4$} & \multirow{2}{*}{$16$} & \multirow{2}{*}{$(7980, 1995)$}                             & \multirow{2}{*}{$665$}                          & $\text{STS}(9)$     & $0, 468, 99, 3, 2, 305, 43, 9, 251, 3, 2$ \\
                         &                       &                                                             &                                                 & $(9 \times 12)$     & $0, 351, 41, 6, 8, 215, 18, 1, 79, 1, 8$ \\ \shhline
    \multirow{4}{*}{$5$} & \multirow{4}{*}{$16$} &                                                             & \multirow{4}{*}{$2562$}                         &                     & $937, 0, 1551, 1264, 1670, 2119, 1973, 1960, 1848, 1223, 1806, \ldots$     \\
                         &                       & $(51240, 20496)$                                            &                                                 & $\text{S-STS}(13)$  & $\quad 15, 1761, 1, 2, 2175, 1169, 1768, 548$ \\
                         &                       & \scriptsize \big($(62500, 25002)$\cite{Esmaeili2010}\big)   &                                                 & $(12 \times 20)$    & $0, 0, 0, 2367, 2491, 126, 2296, 66, 1197, 582, 2200, 9, \ldots$ \\
                         &                       &                                                             &                                                 &                     & $\quad 1836, 2, 1, 0, 1757, 1833, 4$ \\ \shhline
    \multirow{4}{*}{$6$} & \multirow{4}{*}{$16$} &                                                             &                                                 &                     & $8328, 0, 8393, 8106, 7840, 8289, 8143, 8130, 6821, 7393, 6779, 15, 7931, \ldots$ \\
                         &                       & $(227032, 113516)$                                          & $8732$                                          & $\text{STS}(13)$    & $\quad 1, 2, 8345, 7339, 6741, 7390, 1557, 498, 6357, 5666, 5001, 1684$ \\
                         &                       & \scriptsize \big($(229476, 114740)$\cite{Esmaeili2010}\big) & \scriptsize \big($8826$\cite{Esmaeili2010}\big) & $(13 \times 26)$    & $0, 0, 0, 8537, 8661, 126, 8466, 66, 7367, 7424, 8370, 9, 8006, 2, 1, 0,  \ldots$\\
                         &                       &                                                             &                                                 &                     & $\quad 7927, 8003, 4, 0, 8412, 5799, 4553, 2142, 6293$ \\ \shhline
    \multirow{2}{*}{$4$} & \multirow{2}{*}{$18$} & \multirow{2}{*}{$(32676, 8169)$}                            & $2723$                                          & $\text{STS}(9)$     & $0, 853, 217, 6, 2, 1108, 75, 20, 586, 1, 5$ \\
                         &                       &                                                             & \scriptsize \big($2855$\cite{Esmaeili2010}\big) & $(9 \times 12)$     & $0, 1797, 97, 3, 4, 485, 33, 37, 246, 1, 5$ \\ \shhline
    \multirow{4}{*}{$5$} & \multirow{4}{*}{$18$} &                                                             & \multirow{4}{*}{$13588$}                        &                     & $10484, 0, 12275, 10611, 9703, 10786, 10227, 11122, 3263, 7933, \ldots$ \\
                         &                       & $(271760, 108704)$                                          &                                                 & $\text{S-STS}(13)$  & $\quad 3129, 21, 9554, 1, 2, 12183, 7837, 3084, 8297$ \\
                         &                       & \scriptsize \big($(371100, 92777)$\cite{Esmaeili2010}\big)  &                                                 & $(12 \times 20)$    & $0, 0, 0, 12012, 13041, 498, 12534, 223, 7947, 8356, \ldots$ \\ 
                         &                       &                                                             &                                                 &                     & $\quad 12213, 13, 10701, 2, 1, 0, 9550, 10698, 4$ \\ \shhline[1pt]

  \end{tabular}
\end{table*}

\subsection*{Case 2-I: Steiner Triple Systems}
When searching for QC $(J=3, K)$-regular LDPC block codes with girth $g=14,16$, and $18$, we started with a (shortened) base graph constructed by using Steiner triple systems of order $n$, that is, $\text{STS}(n)$ \cite{Johnson2001, Johnson2001-2, Thorpe2004}.

For all $n$, where $n \mod 6$ is equal to $1$ or $3$, there exists a Steiner triple system of order $n$. Then we construct a $(J, K)$-regular, $(c-b) \times c$ base matrix $B$ with entries $b_{ij}$, $i=1,2,\ldots,c-b$ and $j=1,2,\ldots,c$, in such a way that the positions of the nonzero entries in each column correspond to a Steiner triple system of order $(c-b)$. Denote such a $(c-b) \times c$ base matrix $B_{\text{STS}(c-b)}$.

Using the obtained $(J, K)$-regular $(c-b) \times c$ base matrix $B$, we search for a set of edge labels, such that the corresponding voltage graph has at least girth $g$. 

In general, it is possible, without loss of generality, to label a certain subset of edges of the voltage graph simultaneously with zero voltage and thereby decreasing the number of possible labelings. The following algorithm constructs a $(c-b) \times c$ base matrix $B$ based on $\text{STS}(c-b)$ and reorders the matrix to maximize the number of zero entries in its lower left corner. Using such a base matrix, it is always possible to label the last nonzero entry in each column with degree zero. Moreover, in each of the remaining rows at the top of the base matrix, we can label at least one nonzero entry with degree zero.
(Hereinafter we will always choose at least the first element in the remaining rows to be labeled with zero voltage).

\begin{algorithm}[~STS]{Construction of a $(J, K)$-regular $(c-b) \times c$ base graph $B$ obtained from $\text{STS}(c-b)$}
  \begin{enumerate}
    \item Initialize a counter $u$ to zero and denote the current row and column by $s$ and $t$, respectively, starting from the right-most entry in the last row, that is, $s=c-b$ and $t=c$.
    \item Set the $K-u$ elements in row $s$ and column $t,t-1,\ldots,t-K+u+1$ to one, that is, $b_{ij} = 1$ for $i=s$ and $j=t,t-1,\ldots,t-K+u+1$.
    \item Choose the remaining $J-1$ nonzero positions in each of those $K-u$ columns to fulfill the properties of a Steiner Triple System. If possible, choose the positions $b_{ij}$ to minimize $i$. In other words, try to avoid using the lowest rows $s-1, s-2,\ldots$, if possible, despite of the restrictions imposed by the Steiner Triple System.
    \item Finally, decrease $t$ by $K-u$, set $s$ to $s-1$, denote the number of nonzero elements in the new row $s$ by $K-u$ and continue with Step $2$ until all $c$ columns are used, that is, $t=0$.
  \end{enumerate}
\end{algorithm}

By removing the last row and last $K$ columns of the $(J, K)$-regular $(c-b) \times c$ base matrix $B$ constructed using $\text{STS}(c-b)$, we obtain a shortened $(c-b-1) \times (c-K)$ ($J, K-1$)-regular base matrix $B^\prime$, which we denote $B_{\text{S-STS}(c-b)}$. By deleting columns and rows, it is also possible to obtain intermediate codes, which are, however, irregular.

\begin{example}\label{ex: basests}
  In the following we shall construct the $(J=3, K$) base matrices $B$ of dimension $9 \times 12$ $(K=4)$, dimension $13 \times 26$ $(K=6)$ and dimension $25 \times 100$ $(K=7)$. Using Algorithm STS, we obtain the following Steiner Triple Systems of order $9$ ($\text{STS}(9)$), $13$ ($\text{STS}(13)$) and $25$ ($\text{STS}(25)$).
  \begin{IEEEeqnarray*}{rLLLL}
    \IEEEeqnarraymulticol{5}{l}{\text{\textbf{STS}}(9) =} \\
     \Big\{ & \{2, 3, 5\}, & \{1, 4, 6\}, & \{1, 3, 7\}, & \{2, 6, 7\},\\[.1em]
            & \{4, 5, 7\}, & \{1, 2, 8\}, & \{5, 6, 8\}, & \{3, 4, 8\},\\[.1em]
            & \{1, 5, 9\}, & \{2, 4, 9\}, & \{3, 6, 9\}, & \{7, 8, 9\} \Big\} \\[1.2em]
    \IEEEeqnarraymulticol{5}{l}{\text{\textbf{STS}}(13) =} \\
    \Big\{ & \{0, 3, 6\},    & \{0, 2, 7\},  & \{1, 5, 7\},  & \{3, 4, 7\},  \\[.1em]
           & \{3, 5, 8\},    & \{1, 4, 8\},  & \{2, 6, 8\},  & \{2, 4, 9\},  \\[.1em]
					 & \{5, 6, 9\},    & \{0, 1, 9\},  & \{1, 3, 10\}, & \{0, 4, 10\}, \\[.1em]
           & \{6, 7, 10\},   & \{2, 5, 10\}, & \{8, 9, 10\}, & \{7, 8, 11\}, \\[.1em]
           & \{4, 6, 11\},   & \{1, 2, 11\}, & \{0, 5, 11\}, & \{3, 9, 11\},
				\end{IEEEeqnarray*}
				\begin{IEEEeqnarray*}{rLLLL}
           & \{10, 11, 12\}, & \{7, 9, 12\}, & \{0, 8, 12\}, & \{1, 6, 12\}, \\[.1em]
           & \{4, 5, 12\},   & \{2, 3, 12\} \Big\}  \\[1.2em]
    \IEEEeqnarraymulticol{5}{l}{\text{\textbf{STS}}(25) =} \\
    \Big\{ &  \{ 4, 5,10\}, & \{ 1, 9,10\}, & \{ 7, 8,11\}, & \{ 1, 6,11\}, \\[.1em]
           &  \{ 2, 3,12\}, & \{ 0, 9,12\}, & \{ 6, 8,12\}, & \{ 8, 9,13\}, \\[.1em]
			     &  \{ 6, 7,13\}, & \{ 0, 5,13\}, & \{ 2,10,13\}, & \{ 3, 4,14\}, \\[.1em]
			     &  \{ 1,12,14\}, & \{ 0, 2,14\}, & \{ 7, 9,14\}, & \{ 5,11,14\}, \\[.1em]
			     &  \{ 5, 6,15\}, & \{ 3,10,15\}, & \{ 4,12,15\}, & \{ 1, 7,15\}, \\[.1em]
				   &  \{ 0, 8,15\}, & \{11,13,16\}, & \{ 5, 7,16\}, & \{ 6,10,16\}, \\[.1em]
			     &  \{ 2, 8,16\}, & \{ 3, 9,16\}, & \{ 0, 4,16\}, & \{ 9,11,17\}, \\[.1em]
			     &  \{12,13,17\}, & \{ 1, 3,17\}, & \{ 4, 7,17\}, & \{ 0, 6,17\}, \\[.1em]
				   &  \{ 2, 5,17\}, & \{ 8,17,18\}, & \{ 3,11,18\}, & \{ 2, 4,18\}, \\[.1em]
			     &  \{13,15,18\}, & \{ 0,10,18\}, & \{ 1,16,18\}, & \{ 6,14,18\}, \\[.1em]
				   & \{ 9,18,19\}, & \{ 4, 8,19\}, & \{14,15,19\}, & \{10,11,19\}, \\[.1em]
			     & \{ 0, 3,19\}, & \{ 2, 7,19\}, & \{12,16,19\}, & \{ 1, 5,19\}, \\[.1em]
			     & \{17,19,20\}, & \{ 9,15,20\}, & \{10,12,20\}, & \{ 0,11,20\}, \\[.1em]
					 & \{ 5, 8,20\}, & \{ 1, 4,20\}, & \{13,14,20\}, & \{ 3, 7,20\}, \\[.1em]
					 & \{ 2, 6,20\}, & \{ 5,18,21\}, & \{ 4, 6,21\}, & \{ 1,13,21\}, \\[.1em]
			     & \{16,17,21\}, & \{10,14,21\}, & \{ 2, 9,21\}, & \{ 3, 8,21\}, \\[.1em]
           & \{11,15,21\}, & \{ 7,12,21\}, & \{19,21,22\}, & \{18,20,22\}, \\[.1em]
           & \{ 0, 7,22\}, & \{10,17,22\}, & \{ 3, 5,22\}, & \{ 6, 9,22\}, 
	\end{IEEEeqnarray*}
	\begin{IEEEeqnarray*}{rLLLL}
		& \{ 2,15,22\}, & \{ 1, 8,22\}, & \{11,12,22\}, & \{ 4,13,22\}, \\[.1em]
    & \{14,16,22\}, & \{20,21,23\}, & \{ 0, 1,23\}, & \{ 6,19,23\}, \\[.1em]
    & \{15,16,23\}, & \{ 2,11,23\}, & \{ 7,18,23\}, & \{ 5,12,23\}, \\[.1em]
    & \{14,17,23\}, & \{ 4, 9,23\}, & \{ 8,10,23\}, & \{ 3,13,23\}, \\[.1em]
    & \{ 0,21,24\}, & \{22,23,24\}, & \{ 1, 2,24\}, & \{16,20,24\}, \\[.1em]
    & \{ 7,10,24\}, & \{ 8,14,24\}, & \{13,19,24\}, & \{ 3, 6,24\}, \\
    & \{12,18,24\}, & \{15,17,24\}, & \{ 5, 9,24\}, & \{ 4,11,24\}\Big\}  
  \end{IEEEeqnarray*}
  Each number $1,2,\ldots,J$ occurs $K$ times within the set of Steiner triples. However, the chosen Steiner triples are not uniquely determined.

  The corresponding base matrices of dimension $9 \times 12$ $\text{STS}(9)$, dimension $13 \times 26$ $\text{STS}(13)$, and dimension $25 \times 100$ $\text{STS}(25)$ are sparse matrices with nonzero elements only in column $i$ and row $j$, where the $i$th Steiner Triple contains the value $j$. The $9 \times 12$ $(3, 4)$-regular base matrix constructed from $\text{STS}(9)$ denoted by $B_{\text{STS}(9)}$ is given, for example, by
  
  {\setlength{\arraycolsep}{3.5pt}
  \begin{IEEEeqnarray}{rCl}
    \label{eq: base_sts9}    
    B_{\text{STS(9)}} & = & \borderarray{(}{)}{2em}{1.5ex}{ccccccccccccc}{
      ~~~~ & \sm{1}&\sm{2}&\sm{3}&\multicolumn{1}{c}{\sm{4}}&\sm{5}&\sm{6}&\sm{7}&\sm{8}&\sm{9}&\sm{10}&\sm{11}&\sm{12}\vspace{1mm}\\ 
     \sm{1}& 0 & \bs{1} & 1 & 0 & 0 & 1 & 0 & 0 & 1 & 0 & 0 & 0 \\  
     \sm{2}& \bs{1} & 0 & 0 & 1 & 0 & 1 & 0 & 0 & 0 & 1 & 0 & 0 \\  
     \sm{3}& \bs{1} & 0 & 1 & 0 & 0 & 0 & 0 & 1 & 0 & 0 & 1 & 0 \\  
     \sm{4}& 0 & \bs{1} & 0 & 0 & 1 & 0 & 0 & 1 & 0 & 1 & 0 & 0 \\  
     \sm{5}& \bs{1} & 0 & 0 & 0 & 1 & 0 & 1 & 0 & 1 & 0 & 0 & 0 \\  
     \sm{6}& 0 & \bs{1} & 0 & 1 & 0 & 0 & 1 & 0 & 0 & 0 & 1 & 0 \\  
     \sm{7}& 0 & 0 & \bs{1} & \bs{1} & \bs{1} & 0 & 0 & 0 & 0 & 0 & 0 & 1 \\  
     \sm{8}& 0 & 0 & 0 & 0 & 0 & \bs{1} & \bs{1} & \bs{1} & 0 & 0 & 0 & 1 \\  
     \sm{9}& 0 & 0 & 0 & 0 & 0 & 0 & 0 & 0 & \bs{1} & \bs{1} & \bs{1} & \bs{1}
    }\IEEEeqnarraynumspace
  \end{IEEEeqnarray}}
  Entries that correspond to edges in the base graph that can be, according to Algorithm STS, labeled with zero voltage are marked in bold.
  
  By removing the last row and the last $K=4$ columns, the corresponding shortened $8 \times 8$ $(3, 3)$-regular base matrix $B_{\text{S-STS}(9)}$ follows directly as
  {\setlength{\arraycolsep}{3.5pt}
  \begin{IEEEeqnarray}{rCl}
    \label{eq: base_sts9_shortened}
    B_{\text{S-STS(9)}} & = & \borderarray{(}{)}{2em}{1.5ex}{ccccccccc}{
      ~~~~ & \sm{1}&\sm{2}&\sm{3}&\multicolumn{1}{c}{\sm{4}}&\sm{5}&\sm{6}&\sm{7}&\sm{8}\vspace{1mm}\\ 
     \sm{1}& 0 & \bs{1} & 1 & 0 & 0 & 1 & 0 & 0 \\  
     \sm{2}& \bs{1} & 0 & 0 & 1 & 0 & 1 & 0 & 0 \\  
     \sm{3}& \bs{1} & 0 & 1 & 0 & 0 & 0 & 0 & 1 \\  
     \sm{4}& 0 & \bs{1} & 0 & 0 & 1 & 0 & 0 & 1 \\  
     \sm{5}& \bs{1} & 0 & 0 & 0 & 1 & 0 & 1 & 0 \\  
     \sm{6}& 0 & \bs{1} & 0 & 1 & 0 & 0 & 1 & 0 \\  
     \sm{7}& 0 & 0 & \bs{1} & \bs{1} & \bs{1} & 0 & 0 & 0 \\ 
     \sm{8}& 0 & 0 & 0 & 0 & 0 & \bs{1} & \bs{1} & \bs{1}    
    }
  \end{IEEEeqnarray}}
  This corresponds to removing the four Steiner Triples of $\text{STS}(9)$ containing the number of the last row. Shortening the $9 \times 12$ base matrix $B_{\text{STS}(9)}$ constructed from $\text{STS}(9)$ to obtain a shortened $8 \times 8$ base matrix $B_{\text{S-STS}(9)}$ is unpractical as its code rate is $R=1-8/8=0$. However, by shortening the $13 \times 25$ base matrix $B_{\text{STS}(13)}$ in the same way we obtain a $12 \times 20$ base matrix $B_{\text{S-STS}(13)}$ with the feasible code rate $R=8/20$.
\end{example}

In \tabref{girthgreater} the obtained QC ($J=3, K$)-regular LDPC block codes with girth $g=14,16$, and $18$ constructed from Steiner Triple Systems are presented. While the number of nonzero elements in each column is fixed to $J=3$, the number of nonzero elements in each row $K$ is specified in the first column. The second column corresponds to the obtained girth $g$, while in the third and forth columns we give the dimensions of the block code $(n, k)$ after tailbiting to length $M$. And the fifth column contains which Steiner Triple System ($\text{STS}(n)$) is used.

Finally, in the last column $W^{\prime}$ we give the degrees of the corresponding degree matrix $W$ in a compact way. As we have constructed the base matrices in such a way that the last nonzero entry in each column and the first entry in all other rows of the base matrix is labeled with a zero voltage, these entries are omitted. An entry of $W^{\prime}$ in column $j$ and row $i$ corresponds to the voltage degree of the $(j+1)$th nonzero entry in the $i$th row of the corresponding base matrix.

\begin{table}[t]
  \centering
  \caption{\label{tab: girth20}Properties of QC LDPC codes with girth $g \geq 20$}
  \renewcommand\arraystretch{2.2}
  \setlength\tabcolsep{0.4em}
  \begin{tabular}[t]{c!{\shvline[1pt]}c!{\shvline[1pt]}c!{\shvline[1pt]}c!{\shvline[1pt]}c}
    $K$ & $g$  & $(n,k)$             & $M$     & Base graph (\tabref{girth8})\\ \shhline[1pt]
    $4$ & $20$ & $(1296000,      324002)$ &   $36000$ & $(27 \times 36)$,  $g=8$ \\ \shhline
    $5$ & $20$ & $(31200000,   12480002)$ &  $480000$ & $(39 \times 65)$,  $g=8$ \\ \shhline
    $6$ & $20$ & $(518400000, 259200002)$ & $4800000$ & $(54 \times 108)$, $g=8$ \\ \shhline
    $4$ & $22$ & $(7200000,     1800002)$ &  $200000$ & $(27 \times 36)$,  $g=8$ \cite{Esmaeili2010} \\ \shhline
    $5$ & $22$ & $(325000000, 130000002)$ & $5000000$ & $(39 \times 65)$,  $g=8$ \\ \shhline
		$4$ & $24$ & $(39600000,    9900002)$ & $1100000$ & $(27 \times 36)$,  $g=8$ \\ \shhline[1pt]
  \end{tabular}
\end{table}

\subsection*{Case 2-II: $(J, K)$-regular LDPC block codes}
When searching for QC $(J=3, K)$-regular LDPC block codes with girth $g=20, 22$ and $24$, we started with previously obtained QC $(J=3, K)$-regular LDPC block codes of smaller block size and smaller girth, and (re-)applied our algorithms. 

The obtained results for QC $(J=3, K)$-regular LDPC block codes with girth $g=20, 22$ and $24$ are presented in \tabref{girth20}. They are all but one based on previously obtained $(J=3, K)$-regular LDPC block codes with girth $g=8$ (\cf \tabref{girth8}), as specified in their last column in \tabref{girth20}. As before, the first column $K$ denotes the number on nonzero elements in each column; then we give the obtained girth $g$ and the dimensions of the block code $(n, k)$ after tailbiting to length $M$. The corresponding degree matrices are too large and are omitted in \tabref{girth20}, but are available at \cite{online}.

These codes are (probably) unpractical due to their huge block length. However, the table illustrates that by interpreting QC $(J, K)$-regular LDPC block codes as base matrices and re-applying our algorithms we can find QC $(J, K)$-regular LDPC block codes of ``any'' girth $g$.

\section{Conclusions}\label{sec: 9}
Using the relation between the parity-check matrix of QC LDPC block codes and the biadjacency matrix of bipartite graphs, new searching techniques have been presented. Starting from a base graph, a set of edge voltages is used to construct the corresponding voltage graph with a given girth.

By representing bipartite graphs in different ways, lower and upper bounds on the girth as well as on the minimum distance of the corresponding tailbiting block code have been discussed. 

New algorithms for searching iteratively for bipartite graphs with large girth and for determining the minimum distance of the corresponding QC LDPC block code have been presented. Depending on the given girth, the search algorithms are either based on all-ones matrices, Steiner Triple Systems, or QC block codes. Amongst others, new QC regular LDPC block codes with girth between $10$ and $24$ have been presented including their minimum distance if possible. In particular, the previously unknown minimum distance, for some known codes with girth $6$ and $8$, has been determined.

\section*{Acknowledgements}
This research was supported in part by the Swedish Research Council under Grant 621-2007-6281.

\end{document}